\documentclass[12pt]{article}

\usepackage{amsmath,amssymb,euscript,array,cite}

\newcommand{\startappendix}{
\setcounter{section}{0}
\renewcommand{\thesection}{\Alph{section}}}
\newcommand{\Appendix}[1]{
\refstepcounter{section}
\begin{flushleft}
{\large\bf Appendix \thesection: #1}
\end{flushleft}}

\makeatletter
 
  \@addtoreset{equation}{section}
 \makeatother

\def\sst{\scriptscriptstyle}
\def\cms{\widehat{\mathfrak M}}
\def\ms{\mathfrak M}
\def\delbarslash{\,\,{\raise.15ex\hbox{/}\mkern-9mu {\bar\partial}}}
\def\Bomega{\boldsymbol{\omega}}
\def\aD{{\dot\alpha}}
\def\bD{{\dot\beta}}
\def\gD{{\dot\gamma}}
\def\dD{{\dot\delta}}

\def\C{{\bf C}}

\def\R{{\bf R}}
\def\S{{\bf S}}

\def\CB{{\EuScript C}}
\def\N{{\cal N}}
\def\Tr{{\rm Tr}}
\def\CC{{\cal C}}
\def\II{{\mathfrak I}}

\newcommand{\be}{\begin{equation}}
\newcommand{\ee}{\end{equation}}
\newcommand{\bea}{\begin{eqnarray}}
\newcommand{\eea}{\end{eqnarray}}

\newcommand{\MAT}[1]{\begin{pmatrix} #1\end{pmatrix}}
\newcommand{\EQ}[1]{\begin{equation} #1 \end{equation}}

\newcommand{\SP}[1]{\begin{equation}\begin{split} #1
\end{split}\end{equation}}

\begin{document}
\newcommand{\nd}[1]{/\hspace{-0.5em} #1}
\begin{titlepage}
\begin{flushright}
{\bf November 2004} \\ 
hep-th/0411163 \\
\end{flushright}
\begin{centering}
\vspace{.2in}
 {\large {\bf On the Coulomb Branch of a Marginal Deformation} \\
{\bf of ${\cal N}=4$} {\bf SUSY Yang-Mills} 
}\\

\vspace{.3in}
Nick Dorey\\
\vspace{.1 in} 
DAMTP, Centre for Mathematical Sciences \\ 
University of Cambridge, Wilberforce Road \\ 
Cambridge CB3 0WA, UK \\
\vspace{.1 in}
and \\
\vspace{.1 in}
Timothy J. Hollowood \\
\vspace{.1 in}
Department of Physics, University of Wales Swansea \\
Singleton Park, Swansea, SA2 8PP, UK\\
\vspace{.2in}
%
%
\vspace{.4in}
{\bf Abstract} \\

\end{centering}
We determine the exact vacuum structure of a marginal 
deformation of ${\cal N}=4$ SUSY 
Yang-Mills with gauge group $U(N)$. The Coulomb branch of the theory 
consists of several sub-branches which are governed by complex curves of the 
form $\Sigma_{n_{1}}\cup \Sigma_{n_{2}}\cup \Sigma_{n_{3}}$ 
of genus $N=n_{1}+n_{2}+n_{3}$. Each sub-branch intersects 
with a family of Higgs and Confining branches permuted by 
$SL(2,{\bf Z})$ transformations.  We determine the curve 
by solving a related matrix model in the planar limit 
according to the prescription of Dijkgraaf and Vafa, and also 
by explicit instanton 
calculations using a form of localization on the instanton 
moduli space. We find that $\Sigma_n$ coincides with the 
spectral curve of the $n$-body Ruijsenaars-Schneider system. 
Our results imply that the theory on each 
sub-branch is holomorphically equivalent to 
certain five-dimensional gauge theory with eight supercharges.
This equivalence also implies the existence 
of novel confining branches in five dimensions.    

\end{titlepage}
\section{Introduction}
Four dimensional gauge theories 
with ${\cal N}=1$ supersymmetry are of both theoretical and 
phenomenological interest. The vacuum structure of these theories is 
governed by holomorphic observables which can often be computed exactly. 
Unlike other four-dimensional theories, one can determine the spectrum 
of massless particles and the breaking pattern of global and local symmetries 
even at strong coupling. In this paper we will determine the exact 
vacuum structure of an ${\cal N}=1$ model which arises as a marginal 
deformation of ${\cal N}=4$ SUSY Yang-Mills theory with gauge group
$U(N)$ (we also discuss the $SU(N)$ theory).     
  
In terms of ${\cal N}=1$ multiplets, the theory we will consider 
contains a single vector multiplet 
$V$ and three chiral multiplets ${\Phi}_{i}$ $i=1,2,3$ in the adjoint 
representation of the gauge group. The classical superpotential is, 
\begin{equation}
{\cal W}= i\kappa {\rm Tr}_{N}\left[
e^{i\beta/2}\Phi_{1}\Phi_{2}\Phi_{3}-e^{-i\beta/2}\Phi_{1}
\Phi_{3}\Phi_{2}\right]\ .    
\label{LSsupx}
\end{equation}
The ${\cal N}=4$ theory is recovered by setting $\beta=0$ and $\kappa=1$. 
In the following we will refer to the theory with $\beta\neq 0$ as the 
$\beta$-deformed theory.  
This theory is very special as the corresponding deformation of the 
${\cal N}=4$ theory is exactly marginal and gives rise to a 
two-parameter family of ${\cal N}=1$ superconformal field theories
\cite{LS1}.\footnote{For early references on on the UV finiteness of this
 and other four-dimensional models see \cite{early}.}
 
In the absence of scalar vacuum expectation values (VEVs), 
the $\beta$-deformed theory is exactly 
conformally invariant. However the theory also has branches 
of vacua in which conformal invariance is spontaneously broken. 
For generic values of the deformation parameter, 
the theory has Coulomb branches where the gauge group
is broken down to its Cartan subalgebra.  
In addition, for certain special values, we find Higgs branches where the 
gauge group undergoes further breaking. In \cite{nd1}, it was argued
that an exact S-duality, inherited from that of the undeformed 
${\cal N}=4$ theory, implies the existence of dual branches where
magnetic monopoles condense and external electric charges are
confined. This phenomenon is familiar in the context of softly-broken 
${\cal N}=2$ SUSY \cite{SW1} and other ${\cal N}=1$ theories. 
However, there are
some new features in this case. In particular, confinement occurs together
with spontaneously broken conformal invariance, implying the existence
of a massless composite dilaton. As discussed in \cite{nd1,nd2}, the 
large-$N$ physics in these vacua exhibits further novel features
like the appearance of additional dimensions, a relation to 
Little String Theory and a scaling limit where 
the worldsheet theory of the dual string is solvable. 
The new branches occur only
occur at special values of the parameters where the theory is 
strongly-coupled and they are invisible in a classical analysis of
the theory. Part of the motivation for the present investigation is to 
demonstrate the existence of these branches in a more direct way,
without assuming S-duality.    
 
Initially, we will focus on the 
Coulomb branches mentioned above where the low-energy theory has gauge
group $U(1)^{N}$. As in many other examples \cite{SW1,coul}, 
the gauge couplings 
of this abelian effective theory are determined by the period matrix
of a Riemann surface, or curve.
However, the novel feature here is
that the Coulomb branch has the form of the symmetric product
$\text{Sym}_N(\C\oplus\C\oplus\C)$. This means that it 
actually consists of multiple intersecting 
sub-branches which classically can be described by VEVs of the scalar
components of the 3 chiral fields:
\EQ{
\Phi_i=\text{Diag}\big[x^{(i)}_1,\ldots,x^{(i)}_N\big]\ ,
}
where for each $a$ only one of $x^{(i)}_a$ can be non-zero. Each
sub-branch 
$\CB_{n_1,n_2,n_3}$, with $\sum_in_i=N$, is specified by the number
  $n_i$ of $x^{(i)}_a$ that are non-zero. One of our main results is
  that the curve on the sub-branch $\CB_{n_1,n_2,n_3}$ actually has
    disjoint components consisting of three curves of genus $n_i$:
$\Sigma_{n_1,n_2,n_3}=\Sigma_{n_1}\cup\Sigma_{n_2}\cup\Sigma_{n_3}$.
    Furthermore we identify each component $\Sigma_n$ as the spectral
    curve of the $n$-body Ruijsenaars-Schneider (RS) integrable system
    \cite{Ruijsenaars:pv,Ruijsenaars:vq,Ruijsenaars:1986pp}.   

Following earlier work
on massive versions of this model \cite{DHK,Tim, mans}, we determine the
curve (for the simplest case of the branch 
$\CB_{N,0,0}$) using the correspondence between four dimensional 
${\cal N}=1$ theories and zero dimensional matrix models discovered 
by Dijkgraaf and Vafa \cite{DV}. We then  
confirm this result (and extend it to the general case) 
by calculating the instanton contributions
to the couplings of the low-energy effective theory directly. The technical
details of this latter calculation---albeit in a rather condensed form---are 
collected in Appendix B. 
 
Our result for the curve $\Sigma_{n_1,n_2,n_3}$ allows us to identify the
singular points where the Coulomb branch intersects with other Higgs
and confining branches. The singularities occur at points on the
Coulomb branch where states carrying electric and/or magnetic charges
become massless. Although these states are not BPS in an ${\cal N}=1$ 
theory, the monodromies of the period matrix allow us to identify the
quantum numbers of the light states near each singular point. In this
way we can explicitly check the existence of the required set of 
massless magnetic monopoles at the root of the confining branches
discussed in \cite{nd1,nd2}. Our results also provide further
confirmation of the expected $SL(2,{\bf Z})$ duality 
inherited from the undeformed ${\cal N}=4$ theory. In
particular, this duality has a natural action on the moduli space of
the curve $\Sigma_{n_{1},n_{2},n_{3}}$ which permutes the roots of 
Higgs and confining branches.   
        
Remarkably, 
the spectral curve of the $N$-body RS model also governs the
Coulomb branch of another $U(N)$ supersymmetric gauge theory \cite{Nek}. 
This theory lives on the five dimensional spacetime $\R^{3,1}\times
\S^{1}$. Its Lagrangian includes a bare mass term and has eight 
supercharges. Clearly this model is quite different from the
$\beta$-deformed theory, which is four-dimensional, scale invariant
and has only four supercharges. 
Despite these differences, the instanton calculations described in 
Appendix B prove
rather directly that the Coulomb branches of the two theories are 
related by a simple holomorphic 
change of variables. 
We will also see that the two theories have 
Higgs branches of the same dimensions for appropriate 
values of the parameters and they both have $SL(2,{\bf Z})$ 
dualities which act in the same way. 
Our conclusion is that there is a holomorphic equivalence between
the two models. In other words, the two theories are equivalent at the 
level of ${\cal N}=1$ F-terms and differ only by ${\cal N}=1$
D-terms. This equivalence has some interesting
consequences for the five-dimensional theory, which are discussed 
in the final Section of the paper.   
     
The paper is organised as follows. In Section 2, we discuss the
classical vacuum structure of the $\beta$-deformed theory. In Section
3, we investigate the form of perturbative and instanton 
corrections to the classical theory. In Section 4, we apply the 
method of Dijkgraaf and Vafa to obtain the conditions which define the
curve $\Sigma_{N}$ on the branch $\CB_{N,0,0}$. In Section 5, we present a new 
string theory derivation of the complex curve governing the quantum 
Coulomb branch of the five dimensional theory described above. In
particular the string theory construction 
leads to the same defining conditions as those constraining
$\Sigma_{N}$. We then show how      
the spectral curve of the RS system 
produces the unique solution of these conditions. In Section 6, we 
find explicit formulae for the low-energy abelian gauge couplings 
in the case of gauge group $U(2)$ and exhibit the roots of the Higgs
and confining branches. Finally, in Section 7 we discuss some new features 
of the five-dimensional theory which can be inferred from its
holomorphic equivalence to the $\beta$-deformed theory. Some
calculational details are relegated to three Appendices. In
particular, details of the instanton calculations described in the
text are given in Appendix B. After this work
was completed, we recieved the paper \cite{ben}, which has some
overlap with the results presented here.     

\section{Classical vacuum structure}
 
In this Section we will discuss the classical vacuum structure of the 
$\beta$-deformed theory with gauge group $U(N)$. 
The F- and D-flatness conditions are, 
\begin{equation} 
[\Phi_{1},\Phi_{2}]_{\beta}=
[\Phi_{2},\Phi_{3}]_{\beta}=[\Phi_{3},\Phi_{1}]_{\beta}=0
\label{fflat}
\end{equation}
with 
\begin{equation}
[\Phi_{i},\Phi_{j}]_{\beta}=e^{i\beta/2}\Phi_{i}\Phi_{j}-
e^{-i\beta/2}\Phi_{j},\Phi_{i}
\end{equation}
and 
\begin{equation} 
\sum_{i=1}^{3}\,[\Phi_{i},\Phi_{i}^{\dagger}]=0\ ,
\label{dflat}
\end{equation} 
respectively.  
For the ${\cal N}=4$ case, $\beta=0$, 
the deformed commutators appearing 
in the F-term constraint revert to ordinary ones. In this case the vacuum 
equations are solved by diagonalizing each of the three complex scalars, 
\begin{equation}
\langle \Phi_{i} \rangle =  {\rm Diag}\left[ x^{(i)}_{1},
x^{(i)}_{2},\ldots,x^{(i)}_{N}\right]
\label{arb}
\end{equation}
The $3N^{2}$ complex eigenvalues 
$x^{(i)}_{a}$, for $i=1,2,3$ and $a=1,2,\ldots,N$, are unconstrained. 
After taking into account the Weyl group which permutes the eigenvalues, 
we recover 
the familiar Coulomb branch of the ${\cal N}=4$ theory. On this branch the 
$U(N)$ gauge symmetry is spontaneously broken down to its Cartan subalgebra 
$U(1)^{N}$ and the vacuum manifold is the symmetric product 
${\rm Sym}_{N}\, {\bf C}^{3}$.  
 
Introducing a generic, non-zero value of $\beta$ changes things considerably. 
The F-flatness conditions are no longer solved by arbitrary diagonal matrices 
(\ref{arb}). For each value of the Cartan index 
$a\in \{1,2,\ldots,N\}$, at most one of the three eigenvalues,  
$x^{(1)}_{a}$, $x^{(2)}_{a}$ and $x^{(3)}_{a}$, for each $a$, can be 
non-zero. To this end, we define the three subgroups
\EQ{
\II_i=\big\{a\ \big|\ a\in\{1,\ldots,N\}\ ,\ x^{(i)}_a\neq0\big\}\ ,\quad
i=1,2,3\ ;
} 
hence, $\II_1\cup\II_2\cup\II_3=\{1,\ldots,N\}$ and 
$\II_i\cap\II_j=\emptyset$ for $i\neq j$. We also find it convenient
to define for each $a=1,\ldots,N$
\EQ{
x_a=x_a^{(i)}\text{  when  }a\in\II_i\ .
}
In the simplest case of gauge group $U(1)$, 
the Coulomb branch of the 
${\cal N}=4$ theory, which has three complex dimensions, is partially 
lifted leaving three 
complex lines which intersect at the origin. For $G=U(N)$ with $N>1$,  
the Coulomb branch is formed by taking an $N$-fold symmetric product in the 
usual way to give ${\rm Sym}_N(\C\oplus\C\oplus\C)$. 
 So in the $\beta$-deformed theory there are multiple Coulomb branches rather
than a single branch as in an $\N=2$ theory. Up to gauge invariance,
the inequivalent branches are labelled as $\CB_{n_1,n_2,n_3}$, where
$n_i=\text{dim}\,\II_i$.
Often we will pay particular attention to the three branches 
$\CB_1\equiv\CB_{N,0,0}$, etc.,  
on which only one of the one of the three complex adjoint 
scalars is non-zero. 

The Lagrangian corresponding to the 
superpotential (\ref{LSsupx}) has a 
$U(1)^{3}\simeq U(1)^{(1)}_{R}\times U(1)^{(2)}_{R}\times U(1)^{(3)}_{R}$ 
R-symmetry, where each complex 
scalar field $\Phi_{i}$ is charged under $U(1)^{(i)}_{R}$ 
and neutral under the 
other factors. At a generic point on the Coulomb branch 
$\CB_{n_1,n_2,n_3}$ the
complete R-symmetry group is
is spontaneously broken and the gauge group is spontaneously 
broken to $U(1)^{N}$. The branch can also be 
parametrized in terms of the gauge invariant moduli,
\begin{equation}
u^{(i)}_{n}=\frac{1}{N}\langle {\rm Tr}\,\phi_i^{n} \rangle=
\frac{1}{n_i}\sum_{a\in\II_i} \, x_{a}^{n}\ .
\label{moduli}
\end{equation}
for $n=1,2,\ldots,n_i$. 
 
As the unbroken gauge group is $U(1)^{N}$, the massless fields at a generic 
point on the branch includes $N$ massless photons corresponding to 
the diagonal elements
$A_{an}\equiv(A_n)_{aa}$ of the gauge field and 
their gluino superpartners $\lambda_{a\alpha}\equiv(\lambda_\alpha)_{aa}$
which make up $N$ abelian vector multiplets 
of ${\cal N}=1$ SUSY. We denote the corresponding field-strength superfields 
$W_{a\alpha}$, $a=1,2,\ldots ,N$. 
There are also $N$ massless 
chiral multiplets, neutral under $U(1)^{N}$, corresponding to fluctuations 
of the eigenvalues $x_{a}$.
These fields are associated to the subset of
the diagonal elements of $\Phi_i=(\phi_i,\psi_{i\alpha})$; 
namely, $\phi_a=(\phi_i)_{aa}$ and
$\psi_{a\alpha}=(\psi_{i\alpha})_{aa}$ for $a\in\II_i$.
 
In addition to the massless states described above 
the full classical spectrum includes states which have masses due to the 
Higgs mechanism. Such states arise from the non-diagonal elements of
the $U(N)$ vector multiplet
$V$ and also from the non-diagonal plus some of the diagonal elements
of the three chiral multiplets $\Phi_{i}$, $i=1,2,3$. First of all, for
each pair $a\in\II_i$ and $b\in\II_j$, $i\neq j$, there is a vector multiplet
and three chiral multiplets of mass squared $|x_a|^2+|x_b|^2$. For
$a\neq b$ but when $a,b\in\II_i$ there is a vector multiplet
and three chiral multiplets of masses 
$M^{v}_{ab}=|(Z_{v})_{ab}|$ and $M^j_{ab}=|(Z_j)_{ab}|$,
respectively, where
\SP{
(Z_{v})_{ab} &= (Z_i)_{ab} =x_{a}-x_{b}\ ,\\
(Z_{i+1})_{ab} &= e^{i\beta/2}x_{a}-
e^{-i\beta/2}x_{b} \ ,\\     
(Z_{i-1})_{ab} &= e^{-i\beta/2}x_{a}-
e^{i\beta/2}x_{b}\ .
\label{zs}
}
(The labels $i$, etc., are to be understood as defined modulo 3.)
Finally of each $a\in\II_i$ there are 
there are two additional massive chiral multiplets of mass
$M^{i\pm1}_{aa}=2|x_a|\sin(\beta/2)$ coming from the diagonal elements of 
$\Phi_{i\pm1}$.
 
For generic values of the deformation parameter the massless multiplets  
identified above 
correspond to the vanishing diagonal elements $(Z_{v})_{aa}$ and 
$(Z_{i})_{aa}$, $a\in\II_i$.  
However, for special values of the eigenvalues 
$x_a$, additional massless states appear. In some cases, 
these extra massless states indicate 
the freedom to move off along new branches. 
For example, for $a\in\II_i$ there are
always additional massless chiral multiplets $(\Phi_j)_{aa}$, $j\neq
i$ on submanifolds where $x_a=0$. These are points where the three 
different Coulomb branches 
$\CB_{p_1+1,p_2,p_3}$, $\CB_{p_1,p_2+1,p_3}$ and
$\CB_{p_1,p_2,p_3+1}$ intersect. In these cases the 
new massless states are uncharged. As usual there are also 
subspaces on the classical Coulomb branch where a non-abelian subgroup 
of the gauge group is restored. For example, an $SU(2)$ subgroup is restored 
when $x_a=x_b$, for $a,b\in\II_i$. The resulting non-abelian 
low-energy theory is typically asymptotically free and runs to strong coupling 
in the IR invalidating a classical analysis.   

Finally, there are also points where we find additional charged 
massless states without the restoration of non-abelian gauge symmetry. 
In particular, this occurs on the submanifolds where  
$e^{i\beta/2}x_a=e^{-i\beta/2}x_b$, for $a,b\in\II_i$. In these cases the 
resulting low-energy theory is typically IR free and we will below that these 
singular submanifolds persist in the quantum theory. Further 
for certain non-generic rational values of $\beta$, 
the number of massless states is large enough to result in 
a new branch on which these degrees of freedom condense further 
breaking the gauge group.

{\bf\underline{Example: the $U(2)$ theory}}

For simplicity, let us consider the $U(2)$ theory and its
Coulomb branch $\CB_{2,0,0}$
parameterized by eigenvalues $x_{1}$ and $x_{2}$. 
New massless states appear on the one-dimensional submanifolds defined by 
$x_{1}=\exp(\pm i\beta)x_{2}$. The gauge invariant 
version of this condition (for $\beta\neq \pi/2$) is, 
\begin{equation}
u_{1}^{2}=\frac{\cos^{2}({\beta}/{2})}{\cos(\beta)}\,u_{2}\ .
\label{condition}
\end{equation}    
For either root, we find light fields charged  
under $U(1)_{1}\times U(1)_{2}$. For example, near the submanifold  
corresponding to $x_{1}=\exp(-i\beta)x_{2}$, we have two light  
chiral superfields $Q=(\Phi_2)_{12}$ 
and $\tilde{Q}=(\Phi_3)_{21}$ with charges $(+1,-1)$ and $(-1,+1)$. 
This matter content is equivalent to a single ${\cal N}=2$
hypermultiplet and we will sometimes use this language in the following 
(although the theory only has ${\cal N}=1$ SUSY).   
Apart from the gauge couplings, the effective theory of the light fields   
has a superpotential,  
\begin{equation} 
W_{\rm eff}=\left(e^{-i{\beta}/{2}}x_{1}-e^{i{\beta}/{2}}x_{2}\right) 
Q\tilde{Q} 
\end{equation} 
a similar effective theory arises near the other root, corresponding to  
the submanifold $x_{1}=\exp(-i\beta)x_{2}$.  

An important special case is when 
$\beta=\pi$ where the two roots of (\ref{condition}) coincide at  
$u_{1}=0$ or $x_{1}+x_{2}=0$. Near this point in the moduli space  
we now find two light hypermultiplets. Equivalently we find two 
light chiral superfields, $Q_{1}=(\Phi_2)_{12}$ 
and $Q_{2}=(\Phi_3)_{12}$, with charges $(+1,-1)$,  
under $U(1)\times U(1)$ and two more, denoted
$\tilde{Q}_{1}=(\Phi_2)_{21}$ and  
$\tilde{Q}_{2}=(\Phi_3)_{21}$ 
with charges $(-1,+1)$. 
In addition to the gauge couplings,  
the low energy effective theory near this point has superpotential 
\begin{equation} 
W=\left(x_{1}+x_{2}\right)\left(Q_{1}\tilde{Q}_{1}+Q_{2}\tilde{Q}_{2}\right) 
\end{equation} 
A key feature of this effective theory is the existence of a new  
Higgs branch  
on which the massless charged states  
condense. The new branch appears for $x_{1}+x_{2}=0$ 
and allows non-zero values  
for the charged fields subject to the F- and D-term conditions,  
\SP{
Q_{1}\tilde{Q}_{1}+Q_{2}\tilde{Q}_{2}& =0 \\ 
|Q_{1}|^{2}-|\tilde{Q}_{1}|^{2}+|Q_{2}|^{2}-|\tilde{Q}_{2}|^{2} &= 0\ . 
\label{fdc} 
}
By fixing the $U(1)\times U(1)$ gauge symmetry and using the above 
relations, we may  
eliminate $\tilde{Q}_{1}$ and $\tilde{Q}_{2}$, leaving a three complex  
dimensional branch of solutions parametrized by $Q_{1}$, $Q_{2}$ and  
$x_{1}+x_{2}$. When either $Q_{1}$ or $Q_{2}$ is non-zero the  
$U(1)\times U(1)$ gauge symmetry of the Coulomb branch is broken down to  
the diagonal $U(1)$.      

It is not hard to find the corresponding branch in the  
full $\beta$-deformed theory. When $\beta=\pi$, the deformed commutator  
$[\Phi_{i},\Phi_{j}]_{\beta}$ appearing in (\ref{LSsupx})  
becomes the anti-commutator $\{\Phi_{i},\Phi_{j}\}$ and we can solve the  
F- and D-flatness conditions by setting,  
\begin{equation} 
\begin{array}{ccc} \langle \Phi_{1}\rangle = \alpha_{1}\tau_{3} & \qquad  
\qquad  
\langle  \Phi_{2}\rangle = \alpha_{2}\tau_{1} & \qquad \qquad   
\langle \Phi_{3}\rangle = \alpha_{3}\tau_{2} \end{array} 
\label{hb1}
\end{equation}   
wherend $\tau_{i}$ are the Pauli matrices and  
$\alpha_{1}$, $\alpha_{2}$ and $\alpha_{3}$ are arbitrary  
complex numbers. When two or more of the $\alpha_{i}$ are non-zero  
the $U(2)$ gauge group is broken to its central $U(1)$. This  
Higgs branch intersects the Coulomb branch ${\cal C}_{(2,0,0)}$ at $u_{1}=0$  
when $\alpha_{2}=\alpha_{3}=0$. 
The massless modes on the Higgs branch include three scalars corresponding to 
fluctuations of the moduli 
$\alpha_{i}$, $i=1,2,3$ and the photon of the central $U(1)$. 
Each of these bosonic fields is paired with a massless Weyl fermion by 
the unbroken ${\cal N}=1$ supersymmetry. These fields are free at low 
energies and the effective action is precisely that of an ${\cal N}=4$ 
supersymmetric gauge theory with gauge group $U(1)$. The complexified  
gauge coupling of the low-energy theory is related to that of the 
original theory as $\tilde{\tau}=2\tau$. 
 
We will now give a brief (and incomplete) discussion of the  
Higgs branches which appear in  
the $U(N)$ theory for arbitrary $N$. As in the $N=2$ case, the theory   
with $\beta=2\pi/N$ has a Higgs branch  
where $U(N)$ is broken broken to its central $U(1)$.  
The root occurs at a point on the Coulomb branch $\CB_{1}$  
where the eigenvalues of $\Phi_{1}$ take the values  
$x_{a}=\alpha_{1}\exp(2\pi ia/N)$ for $a=1,2,\ldots, N$. As above  
$\alpha_{1}$ is an arbitrary non-zero complex number. At this point  
we find $N$ massless chiral superfields $Q_a$, for $a=1,2,\ldots, N$  
which carry charges,  
\begin{equation} 
\begin{array}{c} 
(+1, -1, 0,\ldots,0) \\ 
(0, +1, -1,\ldots,0)\\  
\ldots\ldots \ldots\ldots \\ 
(0,0\ldots, +1,-1) \\ 
(-1,0,\ldots,0,+1) \end{array} 
\end{equation} 
under the unbroken $U(1)^{N}$ gauge symmetry, as well as charge-conjugate  
degrees of freedom contained in chiral superfields $\tilde{Q}_a$  
with the opposite charges.  The effective theory has also has a  
superpotential which is trilinear in $x_{a}$, $Q_a$ and $\tilde{Q}_a$.   
The matter content and interactions are essentially those of  
an ${\cal N}=2$ quiver theory with gauge group $U(1)^{N}$   
corresponding to the Dynkin diagram of the $A_{N-1}$ Lie algebra.  
The latter theory is known to have a Higgs branch where $U(1)^{N}$ is  
broken to its diagonal $U(1)$ subgroup.  

The corresponding Higgs branch of the full $\beta$-deformed theory  
has scalar expectation values,   
\begin{equation} 
\begin{array}{ccc} \langle \Phi_{1}\rangle = \alpha_{1}U_{(N)} & \qquad  
\qquad  
\langle  \Phi_{2}\rangle = \alpha_{2}V_{(N)} & \qquad \qquad   
\langle \Phi_{3}\rangle = \alpha_{3}W_{(N)} \end{array} 
\end{equation} 
where $U_{(N)}$ and $V_{(N)}$ are the  
$N\times N$ ``clock'' and ``shift'' matrices, $(U_{(N)})_{ab}= 
\delta_{ab}\exp(2\pi ia/N)$ and  
\EQ{
\left(V_{(N)}\right)_{ab}=\begin{cases}1 &
{\rm if}\,\, b=a+1 ,\,\, {\rm mod}\,N \\ 
0 & {\rm otherwise}\end{cases}
\label{shift} 
}
and $W_{N}=V_{(N)}^{\dagger}U_{(N)}^{\dagger}$.  
Here $\alpha_{1}$,  
$\alpha_{2}$ and $\alpha_{3}$ are complex numbers.
     
The $U(N)$ theory can also have more general Higgs branches 
with a larger unbroken gauge group. These occur when the rank $N$ has a 
non-trivial divisor. Thus we have $N=mn$ for some integers $m$ and $n$. 
If the deformation 
parameter takes the value $\beta=2\pi/n$ we find a $3m$ complex parameter 
branch
\EQ{
\langle \Phi_{1}\rangle = \Lambda^{(1)}\otimes 
U_{(n)}\ , \qquad
\langle  \Phi_{2}\rangle = \Lambda^{(2)}\otimes V_{(n)}\ ,\qquad
\langle \Phi_{3}\rangle = \Lambda^{(3)} \otimes W_{(n)}\ ,
\label{vac2}
}
where $\Lambda^{(i)}$ $i=1,2,3$ are three arbitrary 
diagonal $m\times m$ matrices. At a generic point on this branch the 
unbroken gauge symmetry is $U(1)^{m}$. A special case occurs when 
each $\Lambda^{(i)}$ is proportional to the $m\times m$ unit matrix. In this 
three complex parameter subspace the unbroken gauge group is enhanced to 
$U(m)$.           

In addition to the Coulomb and Higgs branches described above,
there are also mixed branches which can be constructed in the obvious
way when $e^{ip\beta}=1$ for $p<N$.

To close this Section we will discuss the case of gauge group 
$SU(N)$. At the classical level, the relation between the $U(N)$ and
$SU(N)$ theories defined by the superpotential (\ref{LSsupx}) is
non-trivial. Apart from the central photon and
its ${\cal N}=1$ superpartner, the $U(N)$ theory also contains three 
chiral superfields $a_{i}={\rm Tr}_{N}\Phi_{i}$ for $i=1,2,3$, which are not
present in the $SU(N)$ theory. While the central $U(1)$ vector multiplet 
is completely decoupled from the $SU(N)$ degrees of freedom, the three 
chiral multiplets do not (for $\beta\neq 0$). All the branches of the $U(N)$ 
theory discussed above are also present in the $SU(N)$ theory,
although, in some cases, their complex dimension is reduced by the 
traceless condition $a_{i}=0$. 

\section{The Quantum Theory}

The superpotential (\ref{LSsupx}) corresponds to an exactly marginal 
deformation of ${\cal N}=4$ SUSY Yang-Mills with gauge group $SU(N)$ 
\cite{LS1}. 
The coupled $\beta$-functions 
for the couplings $\tau$, $\beta$ and $\kappa$ vanish on a two 
(complex) dimensional surface in the parameter space. This surface includes 
the ${\cal N}=4$ line, $\beta=0$, $\kappa=1$ with $\tau$ arbitrary. Away from 
this line the critical surface is specified as 
$\kappa=\kappa_{cr}[\tau,\beta]$ however the explicit form of 
$\kappa_{cr}$ is unknown beyond one-loop. As mentioned above, the 
$U(N)$ theory is classically equivalent to the $SU(N)$ theory with 
additional couplings to the trace chiral multiplets $a_{i}$, for 
$i=1,2,3$. In the quantum theory, these 
additional couplings are actually IR free and thus the 
trace fields decouple from the $SU(N)$ degrees of freedom in the IR
\footnote{The authors acknowledge a useful discussion with Ofer 
Aharony on this point.}. At the quantum level, therefore, 
the $U(N)$ theory at the origin contains 
the $SU(N)$ conformal theory plus some additional free fields.  
               
The vanishing $\beta$-functions imply that the deformed theory has exact 
${\cal N}=1$ superconformal invariance and that there are no chiral anomalies. 
Thus the $U(1)^{3}$ R-symmetry of the classical theory persists in the 
full quantum theory although it is broken spontaneously on the Coulomb
branches. These symmetries prevent the 
Coulomb branch from being lifted by quantum
effects.\footnote{Essentially, a term with the correct R-charge and 
dimension in the effective superpotential would have to be proportional to 
$\Phi_1\Phi_2\Phi_3$ with some contraction of the group indices. 
No such term would lift the Coulomb branch.} Thus on a given Coulomb branch,
we find a moduli space of vacua with $N$ 
massless $U(1)$ vector multiplets
$W_{a\alpha}=(\lambda_{a\alpha},A_{an})$, 
which we can think of
the the diagonal components of the the vector multiplet of the
microscopic theory, and $N$ neutral chiral 
multiplets $\Phi_a=(\phi_a,\psi_{a\alpha})$ 
which we can think of as the diagonal elements 
$(\Phi_i)_{aa}$ for $a\in\II_i$. Although no 
superpotential can be generated, we expect the kinetic 
terms of the massless fields to recieve quantum corrections. As we 
only have ${\cal N}=1$ SUSY the kinetic terms for the scalars correspond to 
D-terms which are relatively unconstrained. In contrast the exact 
effective action for the massless gauge fields is an F-term of the 
form
\begin{equation}
{\cal L}_{\rm eff}=\frac{1}{8\pi}\, {\rm Im}\, 
\left[\int\, d^{2}\theta \, \sum_{ab=1}^N\tau_{ab}(\Phi_c)\, 
W_{\alpha}^{\,a}W^{\alpha b}\,\right]\ .
\label{lefff}
\end{equation}
The effective gauge couplings and vacuum angles are encoded in the 
complex $N\times N$ matrix $\tau_{ab}$, which depends holomorphically 
on the $N$ effective chiral superfields $\Phi_a$, $a=1,2,\ldots N$. In
particular, the coupling constants of the abelian gauge fields at a
point on the Coulomb branch are $\tau_{ab}(x_c)$.

At the classical level the matrix of effective couplings is simply 
$\tau^{\rm cl}_{ab}=\delta_{ab}\tau$. Holomorphy constrains the
possible quantum corrections precisely as in an $\N=2$ theory. 
At the perturbative level, only one-loop
corrections are allowed while beyond perturbation theory 
instanton contributions are allowed of arbitrary charge. These latter
contributions are at leading order in the semi-classical approximation
which is valid for large VEVs. Schematically,
\EQ{
\tau_{ab}(x_{c})=\tau\delta_{ab}+\tau^\text{1-loop}_{ab}(x_{c})+
\sum_{k=1}^\infty\tau^{k-inst.}_{ab}(x_{c})e^{2\pi ik\tau}\ .
\label{instc}
}
In the next subsections, 
we will examine the pertubative and non-perturbative corrections in
turn. 
\subsection{Perturbation theory}

It is straightforward to calculate the effective couplings at one-loop 
in perturbation theory. The result only depends on the masses, spins and
abelian charges of the states that can propagate in the loop. It is
helpful to organize the massive states into multiplets of $\N=1$ supersymmetry.
 
The result is that $\tau_{ab}^\text{1-loop}=0$ for $a=b$ and also for 
$a\in\II_i$ and $b\in\II_j$, when $i\neq j$. For $a,b\in\II_i$ (but $a\neq
b$) we have the non-vanishing contribution
\begin{equation}
\tau_{ab}^{\rm 1-loop} =
f_{ab}-\delta_{ab}\big(\sum_{c\neq a}f_{ac}\big)\ ,
\label{pt1}
\end{equation}
where 
\EQ{
f_{ab}=\frac{i}{2\pi}\log \left[\frac{ 
\left(Z_{v}\right)^{3}_{ab}}{\left(Z_{1}\right)_{ab}\left(Z_{2}\right)_{ab} 
\left(Z_{3}\right)_{ab}}\right]\ .
}
This result reflects the contributions of each of the 
massive states identified in (\ref{zs}) as virtual particles running 
around the loop. 
This formula also exhibits the conformal R-symmetry 
properties of the theory which imply that $\tau_{ab}$ is invariant under the 
transformation $x_{a}\rightarrow \lambda x_{a}$ for $a=1,2,\ldots,N$ 
and $\lambda$ is any complex number.    

Apart from holomorphy, symmetries and the perturbative limit, 
there are other constraints on the exact form of the 
effective gauge couplings. For example, the 
unitarity of the low energy theory 
requires ${\rm Im}[\tau_{ab}]\geq 0$. The low-energy theory on the 
Coulomb branch $\CB_{1}$ is also invariant 
under $Sp(2N,{\bf Z})$ electric-magnetic duality transformations acting on 
the low-energy couplings. This means that, in general, $\tau_{ab}$ will 
not be a single-valued function on the moduli space, but can exhibit 
non-trivial $Sp(2N,{\bf Z})$ monodromies around singular points (or, more 
generally, singular submanifolds). 
Such singular points occur where charged degrees of freedom become massless 
and the monodromies reflect the one-loop $\beta$-function of the 
effective theory of the light degrees of freedom near the singular point.    
An example of this behaviour is already evident in the perturbative result 
(\ref{pt1}). For simplicity we focus on the case $N=2$. 

On the Coulomb branch $\CB_1$ 
of the $U(2)$ theory the gauge symmetry is broken down  
to $U(1)_{1}\times U(1)_{2}$, where the two $U(1)$ factors, generated by  
$Q_{1}$ and $Q_{2}$ respectively, correspond to the two diagonal elements  
of the $U(2)$ gauge field. It is convenient to change basis to  
$U(1)_{\rm even}\times U(1)_{\rm odd}$ generated by $(Q_{1}\pm Q_{2})/2$  
respectively. In terms of the decomposition,  
\begin{equation} 
U(2)\simeq  \frac{U(1)\times SU(2)}{{\bf Z}_{2}} 
\end{equation} 
$U(1)_{\rm even}$ corresponds to the center of $U(2)$ and $U(1)_{\rm odd}$  
corresponds to the Cartan subalgebra of $SU(2)$. They are even and odd 
respectively under the Weyl group of $SU(2)$ which permutes 
$U(1)_{1}$ and $U(1)_{2}$.   

As the gauge boson of $U(1)_{\rm even}$ is decoupled the matrix of abelian  
couplings is diagonal in this basis:  
$\tau_{ab}=
{\rm diag}(\tau_{\rm even},\tau_{\rm odd})$. Including classical and one-loop  
effects we find, $\tau_{\rm even}=\tau$ and  
\begin{equation} 
\tau_{\rm odd}=\tau+\frac{1}{\pi i}\log\left[ 
\frac{(x_{1}-x_{2})^{2}}{ 
(e^{i{\beta}/{2}}x_{1}-e^{-i{\beta}/{2}} 
x_{2})(e^{-i{\beta}/{2}}x_{1}- 
e^{i{\beta}/{2}}x_{2})}\right]\ .
\label{todd} 
\end{equation}          
Clearly this expression exhibits a logarithmic singularity  
on the submanifolds $x_{1}=\exp(\pm i\beta)x_{2}$  
where new massless states appear in the classical theory. It is  
convenient to introduce the gauge-invariant modulus, 
\begin{equation} 
\varphi=\frac{u_{1}}{\sqrt{2u_{1}^{2}-u_{2}}}=\frac{x_{1}+x_{2}} 
{2\sqrt{x_{1}x_{2}}}\ . 
\label{defvarphi}\end{equation} 
The singular submanifolds lie at $\varphi=\pm \cos(\beta/2)$.  
the leading behaviour of $\tau_{\rm odd}$ 
near the point $\varphi=\cos(\beta/2)$ is 
\begin{equation} 
\tau_{\rm odd} \thicksim\,- \frac{1}{\pi i}\log 
\big(\varphi - 
\cos({\beta}/{2})\big)\ . 
\end{equation} 
Thus we see that the effective coupling undergoes a monodromy  
\begin{equation} 
{\cal M}_{1}:\qquad 
\tau_{\rm odd} \rightarrow \tau_{\rm odd} -2\ ,  
\end{equation} 
as we traverse a small circle in the complex $\rho$ plane enclosing  
the point $\varphi=\cos(\beta/2)$ in an anti-clockwise direction.   

In order to have a globally consistent description of the theory with  
${\rm Im}[\tau_{ab}]>0$ everywhere we must find additional singular  
submanifolds with associated $Sp(2N,{\bf Z})$ monodromies which do not  
commute with ${\cal M}_{1}$. These conditions can be satisfied by identifying  
$\tau_{ab}$ with the period matrix of an appropriate family of complex curves  
of genus $N$, just as in an $\N=2$ theory. In the next
section, we will determine this   
curve explicitly. 

\subsection{Instanton effects}

The instanton contributions to the couplings $\tau_{ab}$ in the
low-energy effective action \eqref{instc} 
can be calculated in much the same way as in an $\N=2$ theory. 
Recall that the massless fields correspond to $N$ abelian vector
multiplets of $\N=1$ supersymmetry and $N$ neutral chiral multiplets. The
fermionic components of these multiplets are $\lambda_{a\alpha}$, the
gluinos, and $\psi_{a\alpha}$. (Recall that $\psi_{a\alpha}$ and its
bosonic partner $\psi_a$, for $a\in\II_i$, 
come from the corresponding diagonal component of the
chiral multiplet $\Phi_i$.)
As in an $\N=2$ theory, 
the instanton contribtion to $\tau_{ab}$ can be extracted from the universal 
long distance behaviour of the {\em anti\/}-fermion correlator
\SP{
&\langle\bar\lambda^\aD_a(x^{(1)})\bar\lambda^\bD_b(x^{(2)})
\bar\psi^\gD_c(x^{(3)})
\bar\psi^\dD_d(x^{(4)})\rangle\\
&=\int d^4X\bar S^{\aD\alpha}(x^{(1)}-X)\bar
S^{\bD}_\alpha(x^{(2)}-X)\bar S^{\gD\beta}(x^{(3)}-X)\bar
S^{\dD}_\beta(x^{(4)}-X)\frac{\tau_{ab}}{\partial\phi_c\partial\phi_d}
\Big|_{\phi_a=x_a}+\cdots\
,
\label{afc}
}
where 
\EQ{
\bar S^{\aD\alpha}(x)=\frac1{4\pi^2}\delbarslash^{\aD\alpha}
\Big(\frac1{x^2}\Big)
}
is the free anti-Weyl spinor propagator. 

In order to calculate the instanton contribution to this correlator one
has to insert the leading-order semi-classical expressions 
for the anti-fermions in the instanton background into the measure for
integrating over the supersymmetric multi-instanton moduli space
$\ms_k$.\footnote{We use throughout the notation of \cite{review}.}
Note that in our theory the zero-mode structure is identical to the
$\N=4$ theory and the measure is schematically of the form
\EQ{
\int_{\ms_k}\Bomega^{\sst(\N=4)}e^{-\tilde S^{\sst(\beta)}_k}
}
where $\Bomega^{\sst(\N=4)}$ is the volume form for integrating over
super moduli space of instantons in the $\N=4$ theory. The
$\beta$-deformation appears explicitly in 
$\tilde S_k^{\sst(\beta)}$, the instanton effective action which depends
on the collective coordinates of the instanton. The construction of
this action is outlined in Appendix B. In comparison with the $\N=4$
theory, an instanton configuration only has two supersymmetric
zero-modes. This means with the $\beta$-deformation, the instanton
effective action is only independent of two of the Grassmann
collective coordinates $\xi_\alpha$ associated to the 
supersymmetric zero modes (coming from the gluino). 

In the instanton backgound, the long distance behaviour of the
anti-fermions has the universal form 
\SP{
\bar\psi^\aD_a(x)&=\bar
S^{\aD\alpha}(x-X)\Theta_{\alpha a}+\cdots\ ,\\
\bar\lambda^\aD_a(x)&=\bar S^{\aD\alpha}(x-X)\Xi_{\alpha a}+\cdots\ , 
}
where $\Theta_{\alpha a}$ and $\Xi_{\alpha a}$ are 
functions of the collective coordinates
of the super-instanton that are 
linear in the Grassmann ones. In the above, 
$X$ is the centre of the instanton which is identified with 
the integration variable in \eqref{afc}. The two supersymmetric 
collective
coordinates $\xi_\alpha$ only appear $\Theta_{\alpha a}$ and the
integrals over these variables must therefore be saturated by the two
$\bar\psi$ insertions. The precise form of the relevant terms is
\EQ{
\Theta_{a\alpha}=\xi_\alpha\frac{\partial
\tilde S^{\sst(\beta)}}{\partial x_a}+\cdots\ ,
}  
and therefore 
\EQ{
\tau_{ab}^\text{inst.}=\sum_{k=1}^\infty e^{2\pi ik\tau}
\int_{\cms_k}\Bomega^{\sst(\beta)}e^{-\tilde
  S^{\sst(\beta)}}\,\Xi^\alpha_a\,\Xi_{b\alpha}\ ,
\label{insi}
}
where $\Bomega^{\sst(\beta)}$ is the integral over the $\N=1$ centred
instanton moduli space $\cms_k$ which is, schematically,
$\Bomega^{\sst(\N=4)}/d^4X\,d^2\xi$.

At the one instanton level we can prove that $\tau_{ab}\neq0$ only if
$a,b\in\II_i$. the argument relies on the fact that 
$\Xi_{a\alpha}$ is proportional to one of the Grassmann collective
coordinates associated to the zero-modes of the $\N=4$ theory that are
lifted by the $\beta$-deformations. We denote these as
$\xi^i_\alpha$,  for $i\in\{1,2,3\}$. 
More specifically, for $a\in\II_i$, the exact long distance behaviour
is captured by the exact expression
\EQ{
\Xi_{a\alpha}=\xi^i_\alpha
\frac{\partial\tilde S^{(\beta)}}{\partial x_a}\ .
}
It follows therefore that
\EQ{
\tau_{ab}^\text{1-inst.}
=\frac{\partial^2 F}{\partial x_a\partial x_b}\ ,
\label{ois}
}
for some function $F$. Note that $F$ is like the prepotential of an
$\N=2$ theory except
that in the $\N=1$ context there is no need for \eqref{ois} to be true
for all instanton number. It is given by an integral over the
moduli space quotiented by $\{X_n,\xi_\alpha,\xi^i_\alpha\}$.
It immediately follows
from the existence of $F$ that $\tau_{ab}=0$ for $a\in\II_i$ and
$b\in\II_j$, when $i\neq j$. The reason is that $\tau_{ab}$ is
uncharged under the three R-symmetries $U(1)^{(i)}_R$, $i=1,2,3$. Then, since
$x_a$ has R-charge $+2$ under $U(1)^{(i)}_R$, for $a\in\II_i$, and is
uncharged under the remaining two R-symmetries, it must be that 
$F=F_1+F_2+F_3$, where $F_i$ depends only on $\{x_a,a\in\II_i\}$, 
and $F_i$ must
have the same R-charge as $x_a^2$, $a\in\II_i$. Consequently, $\tau_{ab}$
decomposes into three separate blocks at the one-instanton level. 

It turns out that this block structure generalizes to higher instanton
numbers so that the matrix of couplings is block-diagonal to all
orders in the instanton expansion. 
The proof of this relies on detailed localization arguments
in the instanton calculus and we relegate them to Appendix B. 
So in an exact sense each Coulomb branch $\CB_{n_1,n_2,n_3}$
decomposes into a direct sum of three blocks where the couplings
within each block only depend on the VEVs associated to that block. In
addition, when one of the VEVs, say $x_a$ goes to zero, the
corresponding couplings $\tau_{ab}$ and $\tau_{ba}$, for $b\neq a$ go
to zero. These subspaces describe the intersections of different
Coulomb branches.

\section{The Seiberg-Witten Curve from the Dijkgraaf-Vafa Matrix Integral}
 
The Dijkgraaf-Vafa matrix model gives a way of computing an effective
superpotential in $\N=1$ SYM in terms of the glueball superfields
\cite{DV,DV2,DV3}. In our case, there is a moduli space of vacua and no
superpotential, however, as described originally 
in \cite{Cachazo:2002pr}, the matrix model
technique can be used to re-construct a Coulomb branch by choosing a
suitable deformation which allows one to lift the degeneracy of the
Coulomb branch in a way that allows one to probe an arbitrary
point. This technique is only possible on the Coulomb branches $\CB_i$
where all the VEV reside in one of the chiral fields. For definitness
we choose $\CB_1$.

In order to probe $\CB_1$ we deform the theory 
by a superpotential for $\Phi_1$
of the form $\Tr\,V(\Phi_1)$ where 
\EQ{
V'(x)=\mu\prod_{a=1}^N(x-\xi_a)\ .
}
Classically, the potential lifts $\CB_1$ and leaves an isolated vacuum
at $x_a=\xi_a$.\footnote{There are of course additional vacua, but we
  focus on the one where one eigenvalue of $\Phi_1$ is associated to
  each of the $N$ minima of $V(x)$.} 
We now briefly describe how to apply the matrix
model of Dijkgraaf-Vafa to find the Seiberg-Witten curve on $\CB_1$.

The matrix model involves three 
matrices, one for each of the chiral superfields $\Phi_i$ (see
\cite{DHK,mm1,mm2} for a discussion of these kinds of matrix models).
The matrix model partition function involves
the superpotential of the parent field theory:
\EQ{Z=\int\,\prod_{i=1}^3 d\Phi_i\,\exp-
g_s^{-1}\Tr\left(i\kappa\Phi_1[\Phi_2,\Phi_3]_\beta+
V(\Phi_1)\right)\ .
\label{mm}}
Note that we use the same notation for the matrices as for their
associated chiral superfields. 
The first thing to do 
involves integrating out $\Phi_{2,3}$ by choosing a suitable
contour on which the integrals are well-defined. Then one can go to
eigenvalue basis for $\Phi_1$:
\EQ{
Z\thicksim\int\,\prod_{a=1}^{N}dx_a
\,\frac{\prod_{a\neq b}(x_a-x_b)}
{\prod_{ab}(e^{i\beta/2}x_a-e^{-i\beta/2}x_b)}
\exp -g_s^{-1}\,\sum_a V(x_a)\ .
\label{partfn}
}
Note that the denominator in the above comes from integrating out
$\Phi_2$ and $\Phi_3$ while the numerator is the famous Van der Monde
determinant arising from going to the eigenvalue basis for $\Phi_1$.

Recall that the we want to study the vacuum of the parent theory where
classically $x_a=\xi_a$, $a=1,\ldots,N$. 
In the matrix model, one develops a
saddle-point expand around this classical solution. 
This is achieved by replacing
the size of the matrices by  
$\hat N$ and then by taking $\hat N\to\infty$ with $g_s\to0$ keeping
$S=g_s\hat N$ fixed. In more detail, we take the saddle-point around
the classical critical point with 
$\hat N_a$ eigenvalues at $\xi_a$. Obviously $\sum_{a=1}^N\hat N_a=\hat
N$ and then we take each $\hat N_a\to\infty$ independently keeping the
$N$ quantities
\EQ{
S_a=g_s\hat N_a\ ,\qquad S=\sum_{a=1}^NS_a\ ,
}
fixed. We emphasize that $\hat N_a$ are not the physical degeneracies
which are equal to one.

The saddle-point equation which follows from \eqref{partfn} is
\EQ{
g_s\Big[2\sum_{b(\neq a)} {1\over
{x_a-x_b}}
-\sum_{b} {1\over {x_a-e^{i\beta}
x_b}}-\sum_{b} {1\over {x_a-e^{-i\beta}x_b}}\Big]=
  V^\prime(x_a)\ .\label{spe}
}
The terms on the left-hand side correspond to quantum effects which
modify the classical saddle-point solution. In the large $\hat N$
limit, we can describe the eigenvalues with a density $\rho(x)$ 
in the complex eigenvalue
$x$-plane. Experience suggests that $\rho(x)$ 
has support along $N$ open contours
$\CC_a$ in the neighbourhood of each $\xi_a$. We will
normalize the density via
\EQ{
\sum_a\int_{\CC_a}\rho(x)\,dx=1\ .
}
A key related quantity is the {\it resolvent\/} $\omega(x)$ which is an
analytic function on the $x$-plane with $N$ branch cuts along each $\CC_a$
defined in terms of the density via
\EQ{
\omega(x)=\sum_a\int_{\CC_a} dy \frac{\rho(y)}{x-y}\ .
}
The discontinuity of $\omega(x)$ across a point $x\in\CC_a$ gives the
density:
\EQ{
\omega(x+\epsilon)-\omega(x-\epsilon)
=2\pi i\rho(x)\ ,\qquad x\in\CC_a\ ,
\label{dis}
}
where $\epsilon$ is a suitable infinitesimal.

The saddle-point equation expresses the zero force condition on a test
eigenvalue in the presence of the distribution of a large number 
eigenvalues along each of the cut. It can be written succinctly in
terms of the resolvant as
\EQ{
\frac1{S}V'(x)=2{\bf P}\omega(x)-e^{i\beta}
\omega(e^{i\beta}x)-e^{-i\beta}\omega(e^{-i\beta}x)\ ;\qquad
x\in\CC_a\ ,
\label{nspe}
}
for $a=1,\ldots,N$, 
where ${\bf P}$ implies a principal value, in other words an average of
$\omega(x)$ just above and below the cut at $x$:
\EQ{
{\bf P}\omega(x)=\frac12(\omega(x+\epsilon)+\omega(x-\epsilon))\ ,
}
where $\epsilon$ is a suitable infinitesimal.

The content of this saddle point equation becomes more transparent 
when recast in terms of a new function $t(x)$ (see
also \cite{Hollowood:2004ek}) defined by
\EQ{
t(x)=f(x)+Sx\big(e^{-i\beta}\omega(e^{-i\beta}x)-\omega(x)\big)\ ,
\label{deft}
}
where $f(x)$ is a polynomial defined by
\EQ{
f(x)-f(xe^{i\beta})=xV'(x)\ .
}
>From the analytic
structure of the resolvent $\omega(x)$ it follows that $t(x)$ has cuts along
each $\CC_a$ and its rotation $\CC'_a=e^{i\beta}\CC_a$.
The saddle-point equation \eqref{nspe} is then very simple:
\EQ{
{\bf P}t(x)={\bf P}t(e^{i\beta}x)\ ,\qquad x\in\bigcup_a\CC_a\ .
\label{speta}
}
Given \eqref{dis} this is simply a gluing condition which glues the
top/bottom of $\CC_a$ to the bottom/top of $e^{i\beta}\CC_a$
\EQ{
t(x\pm\epsilon)=t(e^{i\beta}(x\mp\epsilon))\ ,\qquad x\in\bigcup_a\CC_a\ .
\label{spet}
}
So $t$ defines a Riemann surface $\Sigma_N$ of genus $N$ 
which is a copy of the complex $x$-plane with the cuts identified as above.
The function $t$ is then the unique meromorphic 
function on $\Sigma_N$ with a pole at $x=\infty$ of the form
\EQ{
t(x)\underset{|x|\to\infty}\longrightarrow 
f(x)+{\cal O}(1/x)\ .\label{asymp}
}
On the contrary, $x$ is multi-valued on $\Sigma_N$. If we define a basis
of 1-cycles on $\Sigma_N$ $\{A_a,B_a\}$, where $A_a$ encircles $\CC_a$
and $B_a$ joins a point $x\in\CC_a$ to its image $e^{i\beta}x\in
\CC'_a$---and hence is a closed cycle---then $x$ is
single-valued around each $A_a$ but picks up a multiplicative
factor $e^{i\beta}$ around each $B_a$.
It appears that the Riemann surface $\Sigma_N$ has $2N$ complex moduli,
given by the positions of the ends of the $N$ cuts $\CC_a$. However,
the fact that there must exist a function $t$ on $\Sigma_N$ with
the prescribed singularity of order $N$ 
at $x=\infty$, as in \eqref{asymp}, means that
there are actually only $N$ moduli.\footnote{The argument relies on
  the Riemann-Roch Theorem. Firstly, $t$ has a pole of order $N$ 
  at $v=\infty$. There are $2N$ such functions. However
 the singular part of $t$ is fixed \eqref{asymp} giving $N$
  conditions. So the net number of remaining  moduli are $2N-N=N$ as claimed.}

The $N$ moduli of the surface are encoded in the quantities
$S_a=g_s\hat N_a$ which can be expressed as the
following contour integrals:
\EQ{
S_a=S\int_{\CC_a}\rho(x)\,dx=-\frac 1{2\pi
  }\oint_{A_a}\frac{t\,dx}x\ .
\label{shj}
}
The free-energy of the matrix model around the saddle-point solution
$F(S_a)=\log Z$, which is a funciton of the moduli of of the solution
$\{S_a\}$, has the usual topological genus expansion
\EQ{
F(S_a)=\sum_{g=0}^\infty F_g(S_a)g_s^{2g-2}\ .
}
The quantum vacuum of the field theory is described by an effective
superpotential which is a function of the $S_a$ which are now
interpreted as the glueball superfields:
\EQ{
W_{\rm eff}(S_a)=\sum_{a=1}^N
\Big({\partial F_0\over \partial S_a}-2\pi i
\tau S_a\Big)\ ,
\label{gbs}
}
where $\tau$ is the complexified coupling of the
supersymmetric gauge theory in four dimensions and $F_0(S_i)$ is the
genus zero component of the free energy.\footnote{Note that for the
  vacuum in question the eigenvalues are non-degenerate, $N_a=1$. In more
  general vacua the expression for the glueball superpotential
  involves the degeneracy \cite{Hollowood:2004ek}.} 

We already have an expression for the $S_a$ in terms of an integral of
a meromorphic form along 1-cycles of $\Sigma_N$. One can also find a
similar expression for the other quantities in \eqref{gbs}:
\EQ{
\frac{\partial F_0}{\partial
  S_a}=
-i \oint_{B_a}\frac{t\,dx}x\ .
\label{dfq}
}

A critical point of $W_{\rm eff}(S_j)$ corresponds to
\EQ{
\sum_{a=1}^N\frac{\partial^2F_0}{\partial S_b\partial
  S_a}=2\pi i\tau\qquad b=1,\ldots,N\ .
\label{CP}
}
This equation can be written in a more suggestive
way by noticing that
\EQ{
\omega_a=-\frac1{2\pi}\frac{\partial}{\partial S_a}\frac{t\,dx}x\ ,
\quad a=1,\ldots,N
} 
are a basis for the abelian differentials of the first kind on
$\Sigma_N$ normalized by
\EQ{
\oint_{A_a}\omega_b=\delta_{ab}\ .
}
The reason is that the singular part of
$t\,dx/x$ at $x=\infty$ depends only on $V(x)$ and so is manifestly
independent of the moduli $S_a$. Taking the $S_b$ derivative of
\eqref{shj} then proves the result. 
Hence, 
\EQ{ \frac{\partial^2F_0}{\partial S_b\partial
  S_a}=2\pi i\oint_{B_a}\omega_b=2\pi i\tau_{ab}\ ,
}
where $\tau_{ab}$ are elements of the period matrix of $\Sigma_N$.
Consequently the critical point equations are
\EQ{
\sum_{a=1}^N\tau_{ab}=\tau\qquad b=1,\ldots,N\ .
\label{conc}
}
Given that the moduli space of $\Sigma_N$ is 
$N$-dimensional, these $N$ conditions
completely fix the geometry of the Riemann surface $\Sigma_N$
in terms of the parameters of the probe potential $V(x)$.

\subsection{Identification of the critical Riemann surface}

The curve $\Sigma_N$ at the critical point of the glueball
superpotential defines the Seiberg-Witten curve of the $U(N)$ theory
at a point on the Coulomb branch determined by the probe potential
$V(x)$, $x_a=\xi_a$.

The condition \eqref{conc} implies that $\Sigma_N$ at the critical point of
the glueball superpotential is an $N$-fold cover of the torus $E(\tau)$ with
complex structure $\tau$. We can cover $E(\tau)$ with a coordinate $z$
defined modulo $2\omega_1=2\pi i$ and $2\omega_2=2\pi i\tau$ with
$\tau=\omega_2/\omega_1$. The covering map $z(P):\ \Sigma_N\to
E(\tau)$ is then
\EQ{
z(P)=2\pi i
\int_{P_0}^P\sum_{a=1}^N\omega_a\quad\text{mod }2\pi i,2\pi i\tau\ ,
}
where $P_0$ is a fixed, but otherwise arbitrary, base point.

The fact that the curve $\Sigma_N$ is an $N$-fold cover of $E(\tau)$
means that it can
be described by an equation of the form
$F(z,x)=0$ which depends implicitly on the form of $V(x)$; in
other words, on the position on the Coulomb branch $\CB_1$. 
Since $\Sigma_N$ covers $E(\tau)$ $N$ times, the function 
$F(z,x)$ must be of the form
\begin{equation}
F(z,x)=\sum_{a=0}^{N}\, f_a(z)x^a=
\prod_{a=1}^{N}\,(x-x_{a}(z))=0\ ,
\label{fzv}
\end{equation}
where we can choose $f_N(z)=1$. 
The function $F(z,x)$ must satisfy the further conditions:
 
{\bf (i)} Quasi-periodicity in $z$. Notice that $x$ is single-valued
around the $A_a$ cycles but multi-valued around the $B_a$ cycles
(which are lifts of the $A$ and $B$ cycles on $E(\tau)$) we have 
\EQ{
F(z+2\pi i,x)=F(z,x)\ ,\qquad F(z+2\pi i\tau,e^{-i\beta}x)=F(z,x)\ . 
}
In other words, the coefficient functions are quasi-elliptic: 
$f_a(z+2\pi i)=f(z)$ and $f_a(z+2\pi i\tau)=e^{ia\beta}f_a(z)$.
In terms of the $N$ roots $x_{a}(z)$ this condition becomes 
\EQ{
x_{a}(z+2\pi i)=\Sigma^{(1)}_{ab}x_{b}(z)\ ,\qquad
x_{a}(z+2\pi i\tau)=e^{i\beta}\Sigma^{(2)}_{ab}x_{b}(z)\ ,
}
where $\Sigma^{(1)}$ and $\Sigma^{(2)}$ are elements of the permutation 
group $S_{N}$. 
 
{\bf (ii)} Recall that $\Sigma_N$ is a copy of the $x$-plane with cuts
identified in pairs. Hence, $x$ should have a single simple pole on
$\Sigma_N$ corresponding to the point at infinity. Hence,
exactly one of the $N$ roots $x_{a}(z)$ should have a simple pole 
on the torus. Apart from this, the roots $x_{a}(z)$ should have no other 
singularities in the period parallelogram. By choosing a suitable origin for
$z$ we can arrnge this singularity to sit over $z=0$ in the
cover. Consequently $F(z,x)$ behaves near $z=0$ as 
\begin{equation}
F(z,x)\thicksim \frac{g(x)}z +{\cal O}(z^0)\ ,
\end{equation}
where $g(x)$ is a polynomial in $x$ of degree at most $N-1$, and has no 
other singularities in the period parallelogram. 

As mentioned above, condition {\bf (i)} means that the coefficient 
functions $f_{a}(z)$ are quasi-elliptic.  
On the other hand, condition {\bf (ii)}
constrains these functions to have at most a simple pole 
in each period parallelogram. As in the previous Section, one may use an
argument based on the Riemann-Roch
theorem, to count the number of independent 
complex functions satisfying these conditions. 
In this way one finds that the total number of moduli of the 
most general curve satisfying the consitions {\bf (i)} and {\bf (ii)} 
is $N$ as expected. 

\section{Finding the curve}

\subsection{The 5d theory}
 
Our strategy for finding the family of complex curves $\Sigma_N$ which 
satisfies conditions {\bf (i)} 
and {\bf (ii)}, is based on the observation that exactly the same 
conditions arise in the solution of a completely different $U(N)$ gauge 
theory.  The theory in question arises from the compactification 
of a supersymmetric gauge theory in $4+1$ dimensions on a circle 
of radius $R$. The theory has minimal supersymmetry in $4+1$ dimensions 
($8$ supercharges)  
which reduces to ${\cal N}=2$ supersymmetry after compactification to 
$3+1$ dimensions. This theory (to be called simply "the 5d theory" in the 
following) contains a $U(N)$ vector multiplet and a massive 
adjoint hypermultiplet. 
In $4+1$ dimensions the hypermultiplet 
mass is a real parameter $m$. An additional `twisted' mass parameter 
$\mu$ arises after compactification. The twisted mass corresponds to the 
Wilson line of a background gauge field around the compact direction and 
has the periodicity $\mu\sim \mu+2\pi$. The theory has a dimensional 
gauge coupling $G_{5}^{2}$. At energies 
far below the compactification scale the 5d theory reduces to a 
four-dimensional gauge theory with coupling 
$G_{4}^{2}=G_{5}^{2}/2\pi R$. A four dimensional vacuum angle $\Theta$ 
can also be introduced by including appropriate couplings to 
background fields. It is convenient to define 
the complex combination, 
\begin{equation}
M=m+\frac{i\mu}{2\pi R}
\end{equation}
which can be regarded as the lowest component of a  
background vector multiplet of ${\cal N}=2$ SUSY in four dimensions.
 
In $4+1$ dimensions, the vector multiplet contains a real adjoint scalar field 
$\varphi$. The five-dimensional gauge theory described above has a 
Coulomb branch 
parametrized by the eigenvalues of this field. 
An additional adjoint scalar $\omega=\int_{S^{1}}A\cdot dx$ 
arises from the Wilson 
line of the $4+1$-dimensional gauge field around the compact direction. 
The four-dimensional low-energy theory has ${\cal N}=2$ supersymmetry 
and includes a vector multiplet whose lowest component is the complex adjoint 
scalar, 
\begin{equation}
\Phi=\varphi+\frac{i\omega}{2\pi R}
\end{equation}
The Coulomb branch is parametrised in terms of the eigenvalues of 
$\Phi$, 
\begin{equation}
\langle \Phi \rangle 
={\rm diag}(\rho_{1},\rho_{2},\ldots,\rho_{N}) 
\label{eig}
\end{equation}
Apart from the usual action of Weyl gauge transformations 
which permute the eigenevalues, the  periodicity of the Wilson line 
implies the gauge identification: 
$\rho_{a}\sim \rho_{a}+ i/R$ for $a=1,2,\ldots,N$. 
Alternatively we can work in terms of the gauge invariant moduli,  
\begin{equation}
U_{n}=\frac{1}{N}\langle {\rm Tr}_{N}\left[\exp\left(2n\pi R 
\Phi\right)\right] \rangle=
\frac{1}{N}\sum_{a=1}^{N}\exp\left(2n\pi R\rho_{a}\right) 
\end{equation}  
for $n=1,2,\ldots, N$. 
 
The low energy theory on the Coulomb branch is a four dimensional  
$U(1)^{N}$ gauge theory with 
${\cal N}=2$ supersymmetry.  In the quantum theory, the low-energy action 
depends on a matrix $\tau^{\rm 5d}_{ab}$ of complexified abelian couplings 
which  varies as a function of the Coulomb branch moduli $U_{n}$. 
In the exact solution of the system presented in \cite{Nek}, $\tau^{\rm 5d}_{ab}$ 
is identified is identified with the period matrix of the spectral curve 
$\Sigma_{N}$ of the $N$-body Ruijsenaars-Schneider (RS) integrable 
system. We will now explain why this curve naturally provides a 
general solution to conditions equivalent to ${\bf (i)}$ 
and ${\bf (ii)}$. Our strategy will be to realise the classical 5d theory 
on an intersection of branes in Type IIA 
string theory. Following \cite{witm},  we then obtain the quantum 
corrections to the Coulomb branch by lifting to M-theory.  
\subsection{The IIA brane configuration}
 
We begin by considering IIB string theory on $\R^{8,1}\times\S^{1}$ with 
coordinates $x_{0}, x_{1}, \ldots, x_{9}$.  The compact direction 
is parametrized by the coordinate $x_{6}$ with $x_{6}\sim x_{6}+2\pi R_{6}$.  
We introduce $N$  coincident D5 branes with world-volume in the 
$\{0,1,2,3,4,6\}$ directions. At energies far below the string scale 
$M_{s}=1/\sqrt{\alpha'}$, the  worldvolume theory is ${\cal N}=(1,1)$ 
SUSY gauge theory with gauge group $U(N)$ defined on $\R^{4,1}\times\S^{1}$ 
with six-dimensional gauge coupling $G_{6}^{2}=16\pi^{3}\alpha'  g_{s}$. 
At energies below the compactification scale $1/R_{6}\ll M_{s}$, this in
turn reduces to maximally supersymmetric  $U(N)$ gauge theory on
$\R^{4,1}$ with 
gauge coupling $G_{5}^{2}=G_{6}^{2}/2\pi R_{6}$. 
 
We now introduce an additional compact direction via the identification 
$x_{4}\sim x_{4}+2\pi R$ with $R\gg R_{6}$.  The low-energy theory at
scales far below $1/R_{6}$ 
is then the maximally supersymmetric $4+1$ dimensional theory
formulated on 
$\R^{3,1}\times\S^{1}$. 
An equivalent brane configuration which gives rise to the same low energy theory is obtained 
by performing a T-duality transformation in the $x_{4}$ direction. This yields a configuration of 
$N$ D4 branes in Type IIA string theory. The IIA spacetime is 
$\R^{7,1}\times \S^{1}\times \S^{1}$ where 
the compact coordinates are $x_{4}$ and $x_{6}$ with radii $R_{4}=\alpha'/R$ and $R_{6}$. 
The D-branes are wrapped on the $x_{6}$  circle as before but are located at a point in the 
$x_{4}$ direction. 
 
Separating the D4 branes in the compact $x_{4}$ direction corresponds
to turning on a Wilson line for the $(4+1)$-dimensional gauge
field. More generally, the world-volume theory of the D4 branes 
has a moduli space corresponding to the motion of the branes in their transverse directions. 
In terms of minimal supersymmetry in five dimensions (which has
eight supercharges), the theory which lives on the branes includes
a $U(N)$ vector multiplet 
and a single massless hypermultiplet. As mentioned above, the vector mutiplet includes a 
real adjoint scalar $\varphi$ and, after compactification, the Wilson line $\omega$ provides an additional 
scalar. We will focus on configurations where these fields have non-trivial VEVs as described in 
(\ref{eig}) above. These are realised by seperating the D4 branes in the $x_{4}$ and $x_{5}$ directions. 
In particular we define a complex coordinate $u=(ix_{4}+x_{5})/R_{4}$ and place the $D4$ branes at positions 
$u=u_{a}$ for $a=1,2,\ldots, N$ in the complex $u$-plane.  Comparing the spectrum of open 
strings stretched between the D4 branes with the gauge theory spectrum
of W-bosons shows
 that we can identify the positions in terms of the complex 
eigenvalues appearing in (\ref{eig}). Explicitly, we have
$u_{a}=\rho_{a}(2\pi \alpha')/R_4$ for $a=1, 2,\ldots, N$. Note that the 
periodicity of $u$ (which is $u\sim u+ 2\pi i $) 
matches that of the eigenvalues $\rho_{a}$ (namely $\rho\sim\rho+i/R$)
by virtue of the relation $R_{4}=\alpha'/R$. 
 
 Finally, to obtain the 5d theory of interest we must introduce a complex mass $M$ for the adjoint 
 hypermultiplet. This can be accomplished by following the same proceedure used in \cite{witm} 
 to introduce a hypermultiplet mass in the corresponding four-dimensional theory (the 
 ${\cal N}=2^{*}$ theory). To do this we introduce an NS5 brane with world-volume 
 in the $\{0,1,2,3,4,5\}$ directions.  This by itself has no effect on the low energy world-volume theory. 
 To introduce the mass $M$, we include a twist in the boundary conditions in the $x_{6}$ direction. 
 Specifically, rather than simply compactifying the $x_{6}$ direction,
 we divide out by the transformation, 
 \EQ{
 x_{6}\rightarrow x_{6}+2\pi R_{6}\ ,\qquad u \rightarrow u+2\pi RM\ .
 \label{twist}
}
 This twist forces each  D4 brane to break, so that its two  endpoints
 on the NS5 brane are no longer coincident 
 but are seperated by a distance $(2\pi \alpha')|M|$ in the complex $u$-plane. 
 The hypermultiplet degrees of freedom correspond to strings stretched between endpoints 
 of D4 branes on either side of the NS5 and hence they will acquire non-zero masses $|M|$  as 
 required. 

 \subsection{Lifting to M-theory}
  
 To obtain the curve controlling  the Coulomb branch of the 5d theory
 we will lift the IIA brane configuration described above to M-theory.
 Neglecting the twist in the $x_{6}$ direction,  
the IIA spacetime in which the  branes live has the form $\R^{7,1}\times \S^{1}\times \S^{1}$. By IIA/M duality, 
 this is equivalent to M-theory on $\R^{7,1}\times \S^{1}\times \S^{1}\times \S^{1}$ where the additional 
 compact direction, with coordinate $x_{10}$, has radius $R_{10}$. The M-theory parameters 
 are related to the IIA string lengthscale and coupling as, 
 \EQ{
 R_{10}=\sqrt{\alpha'}g_{s}\ , \qquad  M_{\rm
 Pl}=\frac{g_{s}^{-1/3}}{\sqrt{\alpha'}}\ .
}
 Here $M_{\rm Pl}$ denotes the eleven dimensional Planck mass. 
  
 There are two further 
 refinements of the standard IIA/M duality we will need. The first is to introduce a non-trivial vacuum angle 
 $\Theta$ in the low-energy theory on the branes and the second is to reintroduce the 
 hypermultiplet masses via the twist (\ref{twist}). In fact both these modifications can be incorporated 
 by slanting the torus in the M-theory geometry appropriately.  It is convenient to work in terms 
 of dimensionless complex variables $u=(ix_{4}+x_{5})/R_{4}$ and $z=(-x_{6}+ix_{10})/R_{10}$. 
 The resulting M-theory spacetime can be thought of as $\R^{6,1}\times {\cal M}_{\bf C}$ where 
 ${\cal M}_{\bf C}$ is a two dimensional complex manifold with coordinates 
 $u$ and $z$. The complex manifold in question is obtained as a quotient, 
 \begin{equation}
 {\cal M}_{\bf C}= \frac{{\bf C}\times {\bf C}}{\Gamma_{1}\times \Gamma_{2}\times 
 \Gamma_{3}}
 \end{equation}
 where $\Gamma_{1}$,  $\Gamma_{2}$ and $\Gamma_{3}$  are the complex
 translations, 
\SP{
 \Gamma_{1}:\,\,\, z\rightarrow z+ 2\omega_{1} & \qquad \\
 \Gamma_{2}:\,\,\, z\rightarrow z+ 2\omega_{2} & \qquad u\rightarrow
 u+ 2\pi MR\\
 \Gamma_{3}:\,\,\, u\rightarrow u+2\pi i &
}
 with $\omega_{1}=i\pi $ and $\omega_{2}= i\pi \tau$ where 
\begin{equation}
\tau=\frac{iR_{6}}{R_{10}}+ \frac{\Theta}{2\pi}=
\frac{4\pi i}{G_{4}^{2}}+ \frac{\Theta}{2\pi} 
\end{equation}    
In the case $M=0$, dividing out by first two
translations provides the standard definition of the flat complex
torus with complex structure $\tau$, as defined in Section 4.1,
and the manifold ${\cal M}_{\bf C}$ is simply $E(\tau)\times{\bf C}$. 
On reintroducing
a non-zero hypermultiplet mass $M$, this space is no longer a
Cartesian product, but should be thought of as a non-trivial 
complex line bundle over $E(\tau)$ \cite{witm}.
  
Our starting point was a configuration of $N$ D4 branes and a single NS5 brane with intersection 
$\R^{3,1}$. Both types of IIA brane lift to M-theory fivebranes. As usual we expect to find 
a single M5 brane with worldvolume $\R^{3,1}\times \hat{\Sigma}_{N}$
where $\hat{\Sigma}_{N}$ is a Riemann surface 
of genus $N$ embedded as a complex submanifold of ${\cal M}_{\bf
  C}$. As we start from $N$ D4 branes wrapped on the compact $x_{6}$
direction, the resulting M5 brane will wrap the torus $E(\tau)$, N
times. The corresponding Riemann surface will therefore be a branched
$N$-fold cover of $E(\tau)$, which can be described as an $N^\text{th}$ order
polynomial in the variables $z$ and $x=\exp{(-u)}$, 
 \begin{equation}
 F(z,x)=f_{N}(z)\prod_{a=1}^{N}(x-x_{a}(z))
 \end{equation}  
which is automatically invariant under $\Gamma_{3}$. 
For invariance under  $\Gamma_{1}$ and $\Gamma_{2}$ we require, 
\EQ{
F(z+2\omega_{1},x)=F(z,x)\ ,\qquad F(z+2\omega_{2},e^{2\pi
  MR}x)=F(z,x)\ . 
}
This matches precisely with condition ${\bf(i)}$ described in the previous Section provided 
we identify,  
\EQ{
\tau=\frac{4\pi i}{G_{4}^{2}}+ \frac{\Theta}{2\pi}\ ,\qquad
\beta=-2\pi iMR \ .
\label{matching}
}
 
An additional condition on $F(z,x)$ comes from considering the allowed zeros and poles   
of the roots $x_{a}(z)$ on the torus $E(\tau)$. The poles correspond to points on the torus where 
the $M5$ brane goes to infinity in the complex $u$-plane. This is the expected behaviour at points 
where the original IIA brane configuration extends to infinity in the $x_{4}$ and $x_{5}$ directions. 
This occurs only at the position of the NS5 brane, which corresponds
to the point $z=0$. Hence, exactly one of the $N$ roots $x_{a}(z)$ should have a simple pole 
at $z=0$. Apart from this, the roots $x_{a}(z)$ should have no other 
singularities in the period parallelogram. This is equivalent to
condition {\bf(ii)} described in the previous section.
 
In summary the Riemann surface $\hat{\Sigma}_{N}$ controlling the 
Coulomb branch of the 5d theory obeys exactly the same conditions as 
those which constrain the Riemann surface $\Sigma_{N}$ which plays 
the same role for the Coulomb branch $\CB_{1}$ 
in the $\beta$-deformed theory. Fortunately, the Riemann
surface $\hat{\Sigma}_{N}$ is has been determined independently by 
Nekrasov \cite{Nek} to be the spectral curve of the $N$-body 
Ruijsenaars-Schneider (RS) integrable system. We will check
momentarily that the curve indeed provides the unique solution to the
conditions {\bf(i)} and {\bf(ii)} given above. Thus our conclusion
is that the desired complex curve $\Sigma_{N}$ must also be the 
RS spectral curve.     

\subsection{The curve}

The spectral curve of the RS system is given explicitly as, 
\begin{equation}
F(z,x)={\rm det}\left(L(z)-x{\bf 1}_{(N)}\right)=0\ ,
\label{spectral}
\end{equation} 
where $L(z)$ is the $N\times N$ Lax matrix of the RS integrable system with 
elements
\begin{equation}
L_{ab}(z)=i\varrho_{a}\frac{\sigma(q_{ab}-i\beta+z)}{\sigma(x_{ab}-i\beta)
\sigma(z)}e^{\xi(i\pi)\beta z/\pi}\ ,
\label{lax}
\end{equation}
where $q_{ab}=q_{a}-q_{b}$ and 
\begin{equation}
\varrho_{a}=e^{p_{a}}\, \prod_{b(\neq a)} \sqrt{\wp(q_{ab})-\wp(i\beta)}\ .
\label{qa}
\end{equation}
Here $\wp(z)$, $\sigma(z)$ and $\xi(z)$ are standard 
Weierstrass functions for the torus $E(\tau)$.

The first task is to show that $F(z,x)$ 
as defined in (\ref{spectral}) 
satisfies conditions ${\bf (i)}$ and ${\bf (ii)}$ given above in
Section 4.1. 
This is accomplished in Appendix A. We will now 
discuss some of the features of the curve $\Sigma_N$.       
 
The complex parameters $q_{a}$ and $p_{a}$, for $a=1,2,\ldots, N$ 
correspond to the positions and momenta respectively of $N$ particles. 
The spectral curve only depends on these variables through the $N$ conserved 
Hamiltonians; 
\begin{equation}
H_{n}=\sum_{1\leq a_1\leq\cdots\leq a_n\leq N}\prod_{i=1}^n\varrho_{a_i}
\prod_{1\leq i<j\leq n}\frac{1}{\wp(i\beta)-\wp(q_{a_ia_j})}\ ,
\label{hamiltonians} 
\end{equation}
for $n=1,2,\ldots,N$. As explained in the previous 
section this is the expected number of moduli for the most general 
solution of conditions {\bf (i)} and {\bf (ii)}. 
As the Coulomb branch $\CB_1$ 
is parametrized by the $N$ complex moduli $u_{n}$, we should find some 
relation between these quantities and the Hamiltonians $H_{n}$. This 
relation is constrained by the non-anomalous R-symmetry $U(1)^{(1)}_{R}$ 
which acts non-trivially on $\CB_1$. The modulus $u_{n}$ has charge $n$ 
under this symmetry. On the other hand the curve has an obvious symmetry 
under which $v\rightarrow e^{i\alpha}v$, 
$\varrho_{a}\rightarrow e^{i\alpha}\varrho_{a}$, 
under which $H_{n}$ has charge $n$. 
Identifying these symmetries provides a constraint on the relation 
between $u_{n}$ and $H_{n}$ but does not fix it uniquely. For example, 
$u_{2}$ might be identified with any linear combination of $H_{2}$ and 
$H_{1}^{2}$ where the two coefficients can depend on the couplings $\tau$ and 
$\beta$. A comparison with one-loop perturbation theory will at least 
allow us to fix this ambiguity in the weak coupling limit 
$\tau\rightarrow i\infty$. 
 
In fact, the perturbative limit of the period matrix $\tau_{ab}$ of $\Sigma_N$ 
was calculated in the context of the five-dimensional theory described
above \cite{BM1,BM2}. The result reads  
\EQ{
\tau_{ab}\thicksim \delta_{ab}\tau+\frac{i}{2\pi}(1-\delta_{ab})\log 
\left[\frac{\sinh^{2}2\pi R(\rho_{a}-\rho_{b})}
{ \sinh2\pi R(\rho_{a}-\rho_{b}+M)   
 \sinh2\pi R(\rho_{a}-\rho_{b}-M)}\right]\ ,
\label{bm1}
}
where $\rho_{a}$ are the eigenvalues defined in (\ref{eig}) above. The is 
agrees precisely 
with the sum of our classical and one-loop results (\ref{pt1}) provided we 
identify $x_{a}=\exp(2\pi R\rho_{a})$.  
 
In fact we can do much better than this and show by direct calculation 
that the relation between the abelian couplings of the two theories 
described above persists to all orders in the instanton expansion. 
Details of this calculation, which uses localisation techniques to 
calculate the instanton contributions directly are given in Appendix 
B. In summary we find that,   
\EQ{
\tau_{ab}^{(5d)}(\rho_a,M)=\tau_{ab}\big(x_a=e^{2\pi R\rho_a},
\beta=-2i\pi RM\big)\
.
}
which confirms the equivalence of the two Coulomb branch theories 
described above.

The curve $\Sigma_{N}$ has interesting quasi-modular properties under 
$SL(2,{\bf Z})$ 
transformations acting on the microscopic coupling constant $\tau$. 
The modular group acts as, 
\begin{equation}
\tau\rightarrow \tilde{\tau}=\frac{a\tau+b}{c\tau+d}\ ,
\end{equation}
for integers $a$, $b$, $c$ and $d$ with $ad-bc=1$. Under this 
transformation a function $f(z|\tau)$ defined on the torus $E(\tau)$ 
has holomorphic modular weight $w$ if, 
\begin{equation}
f(z|\tilde{\tau})=(c\tau+d)^{w}f(z(c\tau+d)|\tau)\ .
\label{mod}
\end{equation}
With this definition the function the 
Weierstrass function $\wp(z)$ has weight $+2$. The quasi-elliptic functions 
$\xi(z)$ and $\sigma(z)$ have weights $+1$ and $-1$ respectively. 
Thus we find that the equation $F(z,x)=0$ is 
modular invariant if we assign weights $-1$, $0$ to 
$q_{a}$, $\varrho_{a}$ and weights $-1$, $+1$ to $\beta$ and 
$x$ respectively.      

At $\beta=0$, the modular group 
acting on $\tau$ corresponds to the exact S-duality of the ${\cal N}=4$ 
theory. The modular invariance of the curve suggests that S-duality 
extends for non-zero $\beta$, provided we assign a holomorphic 
modular weight of $-1$ to the deformation parameter. Thus S-duality 
transformations relate theories with different values of $\beta$. 
The reason why we should expect such a duality in the 
$\beta$-deformed theory was explained in \cite{DHK,nd1}. At linear 
order, the $\beta$ deformation of the ${\cal N}=4$ theory corresponds to 
adding a (SUSY descendent of a) chiral primary operator $\hat{\cal O}$ 
to the ${\cal N}=4$ Lagrangian with coupling $\beta$. 
As the operator $\hat{\cal O}$ has known modular weight $+1$, modular 
invariance of the ${\cal N}=4$ theory can be restored by assigning the 
coupling $\beta$ weight $-1$

\subsection{The multiple branch structure}

In the earlier parts of Section 5, we have argued that the curve
$\Sigma_N$ describing the Coulomb sub-branches $\CB_i$ is the spectral
curve of the $N$-body RS integrable system. This matches the result of
the instanton analysis. However, the instanton analysis goes much
further in that it describes all the Coulomb sub-branches
$\CB_{n_1,n_2,n_3}$. We can now see how all the sub-branches are
described in the language of the integrable system. First of all,
the roots of the multiple branches occur when one of the eigenvalues
$x_a\to 0$. It is easy to see that this corresponds in the
integrable system to the associated momenta $p_a\to-\infty$ or
$\varrho_a\to0$. In this limit,
the $N$-body RS system naturally degenerates to the $N-1$-body RS
system. Notice that these points of degeneration do not occur in the 
moduli space of the five-dimensional theory since they would require
$\rho_a=-\infty$. As one moves out onto the branch
$\CB_{n_1,n_2,n_3}$, the associated integrable system consists of 3
non-interacting copies of the RS system with $n_1$, $n_2$ and $n_3$
particles. So this Coulomb sub-branch is holomorphically
equivalent to a five-dimensional gauge theory with product gauge group
$U(n_1)\times U(n_2)\times U(n_3)$. This equivalence is confirmed by the 
explicit instanton calculations described in Appendix B.    

\section{Explicit Results for Gauge Group $U(2)$}
 
For $N=2$, the spectral curve of the RS integrable system is, 
\begin{equation}
F(z,x)=x^{2}-H_{1}f_{\beta}(z)x+H_{2}f_{\beta}^{2}(z)\big(\wp(i\beta)-
\wp(z-i\beta)\big)=0\ , 
\label{curven2}
\end{equation}
where
\begin{equation}
f_{\beta}(z)=\frac{\sigma(z-i\beta)}{\sigma(-i\beta)\sigma(z)}e^{
\xi(i\pi)\beta z/\pi}\ .
\label{fz}
\end{equation}
The two Hamiltonians are given as, 
\EQ{
H_{1}=i(e^{p_{1}}+e^{p_{2}})\sqrt{\wp(q_1-q_2)-\wp(i\beta)}\
,\qquad H_{2}=e^{p_{1}+p_{2}}\ .
\label{h2}
}
 
Defining a new variables 
$t=xf_\beta(z)/\sqrt{H_2}-{\cal U}$,  with  ${\cal
  U}=H_{1}/2\sqrt{H_{2}}$, the curve takes on the simpler form, 
\begin{equation}
t^{2}={\cal U}^{2}-\wp(i\beta)+\wp(z-i\beta)\ .
\label{simpler}
\end{equation}
In this form the curve $\Sigma_{2}$ is a 
manifestly a double cover of the standard complex torus
$E(\tau)$ (with 
periods $2\omega_{1}=2\pi i$ and $2\omega_{2}=2\pi i\tau$). Invariance under 
the modular group acting on $\tau$ is manifest if we assign $\beta$ 
holomorphic modular weight $-1$ as above and $t$ and ${\cal U}$ 
both have modular weight $+1$. An interesting double periodicity in $\beta$ 
is also apparant. In particular, note that the theory is obviously invariant 
under shifts of $\beta$ by integer multiples $2\omega_{1}/i=2\pi$. 
This is because the classical superpotential (\ref{LSsupx}) is invariant 
under this shift up to an overall change sign which can be absorbed by redefining the fields. However, the curve is also invariant under shifts of 
$\beta$ by multiples of $2\omega_{2}/i=2\pi\tau$. This periodicity is 
not visible in the classical theory and the period itself is 
non-perturbative in the gauge coupling.         
 
For generic values of $z\in E(\tau)$, the quadratic (\ref{simpler}) 
has two distinct roots
\begin{equation}
t_{\pm}=\pm\sqrt{{\cal U}^{2} -\wp(i\beta)+\wp(z-i\beta)}\ ,
\label{lampm}
\end{equation}  
the branch points of the double cover, occur at special values of $z$ 
for which the two roots coincide: $t_{+}=t_{-}=0$. This occurs 
for values of $z$ satisfying, 
\begin{equation}
\wp(z-i\beta)=\wp(i\beta)-{\cal U}^{2}
\label{branch}
\end{equation}
As $\wp(z)$ is an elliptic function of order two, it attains 
each complex value exactly twice in each period parallelogram. More precisely 
$\wp(z-i\beta)-u$, considered as a function of $z$, 
has exactly two simple zeros for each value of 
$u$, excepting the three special values $u=e_{i}(\tau)$, $i=1,2,3$, for 
which the function has one double zero. Thus, for generic values of 
${\cal U}$, there are two distinct values, $z_{1}$ and $z_{2}$, which    
satisfy (\ref{branch}). These are the two branch points of the 
double-covering, and the two sheets of the covering are joined along a cut 
which runs from $z_{1}$ to $z_{2}$. 
 
As usual we are interested in finding special points in the moduli space 
parametrized by ${\cal U}$, where the curve degenerates. This happens when 
the two branch points $z_{1}$ and $z_{2}$ coincide up to periods of 
$E(\tau)$. From the above discussion, 
this happens only when the RHS of equation (\ref{branch}) attains one of the 
three special values $e_{i}(\tau)$. Thus we find a total of six critical 
points in the moduli space for which
\begin{equation}
{\cal U}={\cal U}_{\pm}^{(i)}=\pm\sqrt{\wp(i\beta)-e_{i}(\tau)}  
\label{crit}
\end{equation} 
for $i=1,2,3$. At these points the curve necessarily degenerates to an 
unbranched double cover of the bare torus $E(\tau)$. 
 
To understand the significance of these points we will compute the period 
matrix $\tau_{ab}$ of the curve which yields the low-energy abelian 
gauge couplings of the $\beta$-deformed $U(2)$ theory. We must first 
find a convenient basis for the homology of $\Sigma$. To do this we define 
a canonical set of cycles on the surface $\{A_a,B_a\}$, with
$a,b=\pm$, and intersections $A_a\cdot B_a=\delta_{ab}$. $A_\pm$
($B_\pm$) are simply the lift of the $A$ and $B$ cycles on the torus
$E(\tau)$ (corrsesponding to $z\sim z+2\pi i$ and $z\sim z+2\pi
i\tau$, respectively) to the two sheets $t=t_\pm$,
where the cycles are chosen to avoid 
the cut in the $z$ plane. 
As we are dealing with a surface of genus two, the space of holomorphic 
differentials is two dimensional. A convenient basis is, 
\EQ{
d\Omega_{1}=dz\ ,\qquad  d\Omega_{2}=dz/t\ .
}
We then define the matrices, 
\EQ{
h_{a\alpha}= \oint_{B_{a}}\, d\Omega_{\alpha}\ ,\qquad e_{a\alpha}= 
\oint_{A_{a}}\, d\Omega_{\alpha}
}
in terms of which the period matrix is computed as the matrix product
\begin{equation}
\tau_{ab}=h_{a\alpha}\left(e^{-1}\right)^{\alpha}{}_{b}
\end{equation}
 
Evaluating this we find, 
\begin{equation}
\tau_{11}=\tau_{22}=\tfrac12(\tau+\tau_\text{odd})\
,\qquad\tau_{12}=\tau_{21}=\tfrac12(\tau-\tau_\text{odd})\ ,
\end{equation}
where $\tau_\text{odd}$ (defined in Section 3.1) is
\begin{equation} 
\tau_\text{odd}=\frac{\oint_{B_{+}}d\Omega_{2}}{\oint_{A_{+}}d\Omega_{2}}
=\frac{\int^{z_0+\omega_{2}}_{z_0}\,\frac{dz}
{  \sqrt{{\cal U}^{2}-\wp(i\beta)+\wp(z-i\beta)}  }}
{\int_{z_0}^{z_0+\omega_{1}}
\,\frac{dz}{ \sqrt{{\cal U}^{2} -\wp(i\beta)+\wp(z-i\beta)}
}}=\frac{-iK'(k)}{K(k)}\ , 
\label{huge}
\end{equation}         
where $z_0$ is arbitrary and 
$K$ and $K'$ are complete elliptic integrals of the 
first kind with parameter
\begin{equation} 
k=\sqrt{\frac{(\wp(i\beta)-
{\cal U}^{2}-e_{1})(e_{2}-e_{3})}{(\wp(i\beta)-
{\cal U}^{2}-e_{2})(e_{1}-e_{3})}}\ .
\label{k1}
\end{equation}
 
The two periods $\tau_\text{even}\equiv\tau$ and $\tau_\text{odd}$ are
even and odd respectively under the ${\bf Z}_{2}$ symmetry
$t\rightarrow -t$ which interchanges the two sheets of
$\Sigma_{2}$. In field theory language, this ${\bf Z}_{2}$ is
precisely the Weyl subgroup of $U(2)$. 
We can compare the extact formula for $\tau_\text{odd}$ 
to the one-loop result \eqref{todd} by taking the 
semiclassical limit $\tau\rightarrow i\infty$. In this limit the 
Weierstrass function reduces to a trigonometric function as, 
\begin{equation} 
\wp(z) \thicksim \frac{1}{4\sinh^{2}\left(\frac{z}{2}\right)}+\frac{1}{12}\ ,
\end{equation} 
while the quasi-modular forms $e_{1}$, $e_{2}$ and $e_{3}$, tend to the 
constant values $-1/6$, $1/12$ and $1/12$ respectively. Using standard results 
for the asymptotics of the complete elliptic integrals we obtain, 
\begin{equation}
\tau_{\rm odd}\thicksim \tau+   \frac{1}{i\pi}\log\left(\frac{4{\cal U}^{2}+
\sin^{-2}(\beta/2)}{1-4{\cal U}^{2}-
\sin^{-2}(\beta/2)}\right)\ .
\label{pert}
\end{equation}
This matches the perturbative result \eqref{todd}, provided that we set 
\begin{equation}
{\cal U}=\frac{H_{1}}{2\sqrt{H_{2}}}=\frac{i\varphi}
{2\sin(\beta/2)}= \frac{i}{4\sin(\beta/2)}
\left(\sqrt{\frac{x_{1}}{x_{2}}}+\sqrt{\frac{x_{2}}
{x_{1}}}\right)\ ,
\label{modulus}
\end{equation}
where $\varphi$ was defined in \eqref{defvarphi}.  

We can also look at the behaviour of ${\tau}_{\rm odd}$ near the six 
points in moduli space where the curve degenerates. Near the points 
${\cal U}^{(1)}_{\pm}$, defined in (\ref{crit}) above, we find, 
\begin{equation}
{\tau}_{\rm odd}\thicksim -\frac{1}{i\pi}\log\left({\cal U}-{\cal U}^{(1)}_{\pm}
\right)+\ldots
\label{u1}
\end{equation}
The coefficient in front of the logarithm is consistent with the 
appearance of a single massless hypermultiplet electrically charged under 
$U(1)_{\rm odd}$. We can confirm this interpretation by noting that, 
in the semiclassical limit $\tau\rightarrow i\infty$, we have, 
${\cal U}^{(1)}_{\pm}\simeq \pm i\cot(\beta/2)/2$. Using the semiclassical 
identification (\ref{modulus}) this corresponds the relations $\lambda_{1}=
\exp(\pm i\beta)\lambda_{2}$ between the two eigenvalues of $\Phi_{1}$. 
These are precisely the points where massless charged hypermultiplets appear 
in the classical theory. 
 
We next consider the neighbourhood critical points,\footnote{For 
pedagogical reasons it is convenient to discuss the
  critical points in the order ${\cal U}^{(1)}_{\pm}$, then 
${\cal U}^{(3)}_{\pm}$ then ${\cal U}^{(2)}_{\pm}$.} ${\cal U}^{(3)}_{\pm}$.
Near either of these points the abelian couplings exhibit the asymptotics,   
\begin{equation}
{\tau}_{\rm odd}\thicksim -\frac{i\pi}{\log\left({\cal U}-{\cal U}^{(3)}_{\pm}
\right)} + \ldots
\label{u3}
\end{equation}
Thus the low-energy gauge-coupling $g^{2}_{\rm odd}=
4\pi/{\rm Im}\tau_{\rm odd}$ has a logarithmic divergence near 
these points. To interpret this we recall the electric-magnetic duality 
of the low-energy action which, for $N=2$ includes an $SL(2,{\bf Z})$ 
acting on $\tau_{\rm odd}$. After performing the duality transformation 
$\tau_{\rm odd}\rightarrow -1/\tau_{\rm odd}$, the asymptotics of the 
dual coupling precisely match those expected for a massless 
hypermultiplet ({\it i.e.\/} (\ref{u1})). The intepretation is therefore that 
we have a massless hypermultiplet at each of the points 
${\cal U}^{(3)}_{\pm}$ which is minimally coupled to the 
dual gauge-field, in other words these degrees of freedom carry magnetic 
charge. 
 
Finally in the vicinity of the third critical point 
${\cal U}^{(2)}_{\pm}$ the behaviour of the coupling is,        
\begin{equation}
{\tau}_{\rm odd}\thicksim -\frac{i\pi}{\log\left({\cal U}-{\cal U}^{(2)}_{\pm}
\right)}\,\,\, -1+\ldots
\label{u2}
\end{equation}
Similar reasoning shows that dyonic hypermultiplets carrying both 
electric and magnetic charges become massless at these points. 
 
In summary, we have found six critical points in the moduli space. 
At each of these points an additional massless hypermultiplet appears. 
The points come in three pairs with massless electric, magnetic and dyonic 
degrees of freedom respectively. As we vary the deformation parameter $\beta$, 
we can find special values at which one pair of these points coincide. 
Here we will focus on the cases where the massless degrees of freedom are 
mutually local. Thus we look for values of $\beta$ for which 
${\cal U}^{(i)}_{+}={\cal U}^{(i)}_{-}=0$ for exactly one value of $i$. 
This requires $\wp(i\beta)=e_{i}(\tau)$. This is satisfied when 
$i\beta$ equals one of the the half periods 
$\omega_{1}$, $\omega_{1}+\omega_{2}$ or $\omega_{2}$. The three special 
values are therefore:
 
{\bf (1)} For $\beta=\pi$ we have $\wp(i\beta)=e_{1}(\tau)$ 
which implies ${\cal U}^{(1)}_{+}={\cal U}^{(1)}_{-}=0$. Thus, for this 
special value of $\beta$, two electrically charged hypermultiplets appear 
at the point ${\cal U}=0$. Near the point ${\cal U}=0$ we therefore
have four light electrically charged chiral multiplets $Q_{i}$ and 
$\tilde{Q}_{i}$ for $i=1,2$. To have these multiplets become 
massless at the point ${\cal U}=0$, we must have a superpotential 
of the form
\begin{equation}
W\thicksim x\left(Q_{1}\tilde{Q}_{1}+
Q_{2}\tilde{Q}_{2}\right)\ ,
\end{equation}
where $x$ is some local coordinate on the Coulomb branch which goes to 
zero at the singular point. As explained in Section 2, such an
effective theory has a Higgs branch on which the charged chiral multiplets
condense breaking the gauge group down to its $U(1)$ center.  
In fact this is just the Higgs branch (\ref{hb1}) which is 
already visible in the classical theory. Our analysis confirms that
the Higgs branch is still present in the quantum theory and 
intersects the Coulomb branch at ${\cal U}=0$ as in the classical theory.  
    
{\bf (2)} For $\beta=\pi\tau$ we have $\wp(i\beta)=e_{3}(\tau)$ 
which implies ${\cal U}^{(3)}_{+}={\cal U}^{(3)}_{-}=0$. For this 
value of $\beta$, the monodromies are consistent with 
the existence of four massless magnetically charged chiral 
multiplets $Q_{i}^{(M)}$ and $\tilde{Q}_{i}^{(M)}$, with $i=1,2$, 
at the point ${\cal U}=0$. In addition to magnetic gauge couplings, 
the effective theory of the light degrees of freedom has a
superpotential term of the form  
\begin{equation}
W\thicksim x^{(M)}\left(Q^{(M)}_{1}\tilde{Q}^{(M)}_{1}+
Q^{(M)}_{2}\tilde{Q}^{(M)}_{2}\right)\ ,
\end{equation}
where $x^{(M)}$ is a local coordinate 
on the Coulomb branch which vanishes at 
the singular point ${\cal U}=0$. 
This effective theory has a three complex dimensional branch on which
the magnetically charged fields have non-zero VEVs. Thus we find a 
branch on which magnetic states 
condense leading to the confinement of electric charges. 
This is one of the confining branches discussed in \cite{nd1}. 
 
{\bf (3)} For $\beta=\pi(\tau+1)$ we have $\wp(i\beta)=e_{2}(\tau)$ 
which implies ${\cal U}^{(2)}_{+}={\cal U}^{(2)}_{-}=0$. 
 states carrying both electric and 
magnetic charges at the point ${\cal U}=0$. The effective theory near the 
singular point has a branch on which these dyonic degrees of freedom 
condense. This theory branch on this branch therefore features 
oblique confinement.  
 
It is interesting to look at the form of the curve which appears at the 
critical values of $\beta$ identified above. We will work in terms of the 
original form of the $U(2)$ curve (\ref{curven2}). In each case 
the critical value of the modulus ${\cal U}=0$ corresponds to $H_{1}=0$, 
with arbitrary $H_{2}$. On this locus the curve becomes; 
\begin{equation}
x^{2}=H_{2}f^{2}_{\beta}(z)
\left(\wp(z-i\beta)-\wp(i\beta)\right)
\label{critcurve}
\end{equation}
At the critical value $\beta=\pi$ corresponding to the root of the Higgs 
branch the curve simplifies dramatically and becomes,  
$x^{2}=H_{2}F_{1}(\tau)$ where $F_{1}(\tau)=2e_{1}^{2}(\tau)+
e_{2}(\tau)e_{3}(\tau)$ is a quasi-modular form of 
weight four. The two roots $x_{\pm}(z)$ are equal to the constants $\pm 
\sqrt{H_{2}F_{1}(\tau)}$. Under translation by the two periods of the 
torus the roots behave as, 
\EQ{
x_{\pm}(z+2\omega_{1})=x_{\pm}(z)\ ,\qquad
x_{\pm}(z+2\omega_{2})=e^{i\beta}x_{\mp}(z)\ .
\label{s1}
}
Thus the minimal half-periods $\tilde{\omega}_{i}$ such that $x_{\pm}(z+
2\tilde{\omega}_{i})=x_{\pm}(z)$ for $i=1,2$ are 
$\tilde{\omega}_{1}=\omega_{1}$ and $\tilde{\omega}_{2}=2\omega_{2}$. 
The corresponding critical curve is an unbranched double cover of the torus 
$E(\tau)$ with modular parameter $\tilde{\tau}=2\tau$.  
 
It is also instructive to look more closely at the behaviour of the curve 
as we approach the critical point. 
Thus we set $\beta=\pi+\epsilon$ and expand (\ref{critcurve}) 
to first order in $\epsilon$, 
\begin{equation}
x^{2}=H_{2}F_{1}(\tau)+ \epsilon{\cal A}(z)
\label{cala} 
\end{equation}
Although the second term is subleading in $\epsilon$, it becomes large 
near the point $z=0$. In fact we find ${\cal A}(z) \sim c/z$ 
plus finite terms as $z\rightarrow 0$, where 
$c=-i\wp''(i\pi)$. Near the origin (\ref{cala}) becomes, 
\begin{equation}  
z(x-x_{+})(x-x_{-})=c\epsilon
\end{equation}
Thus we see that, as $\epsilon\rightarrow 0$, the curve actually 
factorizes into 
two branches. One branch, defined by $x=x_{\pm}$ with arbitrary $z$ 
corresponds to the double cover of the torus $E(\tau)$ discussed above. 
The new branch is defined by $z=0$ with arbitrary $x$. Factorization of 
the curve into two pieces is the usual signal of the appearance of 
a new branch: in this case the Higgs branch discussed above. 
In the next section we will make this observation precise 
in the context of the M-theory construction of the curve. 
 
The behaviour of the curve at the other two critical points 
is related to the behaviour at the Higgs branch root via S-duality.
For $\beta=\pi\tau$ and $H_{1}=0$, the curve becomes 
$x^{2}=H_{3}F_{3}(\tau)\exp(\pi i z/\omega_{1})$ with 
$F_{3}(\tau)=2e_{3}^{2}+e_{1}e_{2}$. 
The two roots are $v_{\pm}(z)=\pm 
\sqrt{H_{2}F_{3}(\tau)}\exp(\pi iz/2\omega_{1})$. 
Under translation by the two periods of the 
torus the roots behave as, 
\EQ{
x_{\pm}(z+2\omega_{1})=x_{\mp}(z)\ ,\qquad
x_{\pm}(z+2\omega_{2})=e^{i\beta}x_{\pm}(z)\ .
\label{s3}
}
Thus the minimal half-periods $\tilde{\omega}_{i}$ such that $x_{\pm}(z+
2\tilde{\omega}_{i})=x_{\pm}(z)$ for $i=1,2$ are 
$\tilde{\omega}_{1}=2\omega_{1}$ and 
$\tilde{\omega}_{2}=\omega_{2}$. The critical curve is an 
therefore an unbranched double cover of the torus 
$E(\tau)$ with modular parameter $\tilde{\tau}=\tau/2$. 
 
Finally, for $\beta=\pi(\tau+1)$ and $H_{1}=0$, the curve becomes 
$x^{2}=H_{2}F_{2}(\tau)\exp(\pi i z/\omega_{1})$ with 
$F_{2}(\tau)=2e_{2}^{2}+e_{1}e_{3}$. 
The two roots are $x_{\pm}(z)=\pm 
\sqrt{H_{2}F_{2}(\tau)}\exp(\pi iz/2\omega_{1})$. 
Under translation by the two periods of the 
torus the roots behave as, 
\EQ{
x_{\pm}(z+2\omega_{1})=x_{\mp}(z)\ ,\qquad
x_{\pm}(z+2\omega_{2})=e^{i\beta}x_{\mp}(z) 
\nonumber \\
\label{s2}
}
Thus the minimal half-periods $\tilde{\omega}_{i}$ such that $x_{\pm}(z+
2\tilde{\omega}_{i})=x_{\pm}(z)$ for $i=1,2$ are 
$\tilde{\omega}_{1}=2\omega_{1}$ and 
$\tilde{\omega}_{2}=\omega_{1}+\omega_{2}$. 
Thus the critical curve is an unbranched double cover of the torus 
$E(\tau)$ with modular parameter $\tilde{\tau}=(\tau+1)/2$.   
 
In each case the resulting surface is interpreted as an unbranched 
double cover of the torus $E(\tau)$ which is itself a torus $E(\tilde{\tau})$ 
with complex 
structure parameter $\tilde{\tau}=\tilde{\omega}_{2}/\tilde{\omega}_{1}$. 
Thus in the three degenerate cases $\beta=\pi$, $\pi\tau$ and 
$\pi(\tau+1)$ we found $\tilde{\tau}=2\tau$, $\tau/2$ and $(\tau+1)/2$ 
respectively. These values correspond to the three inequivalent 
unbranched double covers of $E(\tau)$ and they are naturally permuted 
by S-duality. 
 
The generalisation of these results to gauge group $U(N)$ with $N>2$ is
straightforward. As explained in Section 2, the classical theory has a
Higgs branch on which the unbroken gauge group is $U(1)$ for 
$\beta=2\pi/N$. This branch survives in the quantum theory and
corresponds to a degeneration of $\Sigma_{N}$ to an unbranched
$N$-fold cover of the torus $E(\tau)$ with complex structure
$\tilde{\tau}=N\tau$. In the quantum theory, there are also branches 
in confining phases which are related to the Higgs branch by
S-duality. These correspond to the degenerations of $\Sigma_{N}$ into
inequivalent $N$-fold covers of $E(\tau)$. These correspond to all torii 
with complex structure $\tilde{\tau}=(p\tau+k)/q$ where $pq=N$ and 
$k=0,1,\ldots,q-1$. Thus the total number of inequivalent branches is 
equal to the sum of the divisors on $N$. 
Note that this is essentially identical
to the classification of massive vacua of the ${\cal N}=1^{*}$ theory 
\cite{VW,DW,ND}. In the present case, each branch occurs at a set
of critical values of the form $\beta=2\pi(l+ m\tilde{\tau})/p$ 
where $l$ and $m$ are integers.  

\section{Discussion}
 
An interesting consequence of the analysis given above is that two
very different supersymmetric gauge theories have Coulomb branches 
described by the same complex curve $\Sigma_{N}$. The 
$\beta$-deformed theory, which is the main subject of the paper, lives 
in $3+1$ dimensions, has four supercharges and also has spontaneously broken
conformal invariance. The other theory (the 5d theory of Section 5 above) 
lives in $4+1$ dimensions, has eight supercharges and no conformal
invariance. Despite these differences, our results imply that, on
their Coulomb branches these two models 
agree exactly at the level of ${\cal N}=1$ F-terms. Indeed, below 
Eq (\ref{bm1}), we identified the holomorphic 
change of variables which relates the two Coulomb branch theories at
one loop: it is simply the exponential map $x_{a}=\exp(2\pi R\rho_{a})$ for
$a=1,2,\ldots, N$. In Appendix B, we check that this relation
continues to hold order by order in the instanton expansion and
therefore for the exact matrix of abelian couplings. 
In this Section, we will check that the pattern of Higgs and 
Confining branches in the two theories also agrees. 
Our conclusion therefore is that the theories are actually
holomorphically equivalent in the sense explained in Section 1. 
In this Section we will also discuss some of the consequences of 
this equivalence for the 5d theory. 
 
As the curves for the two theories agree, the Coulomb branches of 
both models have the same singular points, with equivalent monodromies 
and massless states. The 5d theory should therefore exhibit the same 
Higgs and Confining branches intersecting the Coulomb branch at 
the singularities. The Higgs branch roots of the 5d theory are
straightforward to find. As the 5d theory has eight supercharges the
stable states of the theory are BPS saturated. The 
classical mass spectrum of elementary 
quanta on the 5d Coulomb branch is governed by the central charge, 
\begin{equation}
{\cal Z}_{ab}=\frac{ik}{R}+ \rho_{a}-\rho_{b}\pm M
\label{ccharge5d}
\end{equation}      
Here the integer $k$ corresponds to momentum around the compact
$x_{4}$ direction which has radius $R$. The complex eigenvalues
$\rho_{a}$, for $a=1,2,\ldots,N$ are defined in Section 5.1 above as 
is the complex mass $M$. A Higgs branch of the $U(N)$ $\beta$-deformed 
theory occurs at $\beta=2\pi/N$. According to (\ref{matching}), the
equivalent value of the 5d parameter $MR$ is $i/N$. If we also specify 
the Coulomb branch VEVs as $\rho_{a}=a i/NR$ for $a=1,2,\ldots,N$ the 
central charge becomes,  
\begin{equation}
{\cal Z}_{ab}=\frac{i}{NR}\left(kN+ a-b\pm 1\right)
\label{ccharge5d22}
\end{equation}     
Thus we find a total of $N$ massless ${\cal N}=2$ 
hypermultiplets coming from off-diagonal elements of 
${\cal Z}$ with\footnote{Each ${\cal N}=2$ hypermultiplet includes 
two ${\cal N}=1$ chiral multiplets of opposite charges.} 
$a=b\pm 1$ mod $N$. Each of these fields 
carry opposite charges under a pair of adjacent $U(1)$'s 
in the Cartan subalgebra of $U(N)$. This precisely the same 
massless spectrum which appears at the root of the 
corresponding Higgs branch of the $\beta$ deformed theory.   
 
The presence of the Higgs branch in the classical theory can be 
seen directly in the IIA brane construction of Section 5.2. 
The values for the complex eigenvalues $\rho_{a}$ described 
above correspond to a configuration where the $N$ D4 branes are
distributed around the compact $x_{4}$ direction with equal spacings 
$2\pi R_{4}/N$. At the special value $MR=i/N$ the twist (\ref{twist}) in
the $x_{6}$ direction becomes,   
\EQ{
 x_{6}\rightarrow x_{6}+2\pi R_{6}\ ,\qquad x_{4} \rightarrow 
x_{4}+\frac{2\pi R_{4}}{N}\ .
 \label{twist2}
 }
As the resulting shift in $x_{4}$ is equal to the seperation between
branes, the $N$ D4 branes segments form a closed spiral. In other words they
form a single D4 wrapped $N$ times around the $x_{6}$ direction and
once around the $x_{4}$ direction. This D4 can now move away from the 
NS5 in the $x_{7}$, $x_{8}$ and $x_{9}$ directions. 
As we now have a single D4 brane, this corresponds to the expected 
Higgs branch where $U(N)$ is broken down to $U(1)$.  
The D4 can also move 
parallel to the NS5 in the $x_{4}$ and $x_{5}$ directions and we we may 
turn on a Wilson line for the $U(1)$ world-volume gauge field around
the $x_{6}$ circle. Thus the Higgs branch has a total of six real or
three complex dimensions as expected. One may easily check that other 
Higgs branches which occur when multiple spirals can be formed match
the remaining classical Higgs branches of the $\beta$-deformed theory.  
 
The brane picture of the Higgs branch root can easily be lifted to
M-theory. The D4 brane described above lifts to a single M5 with two
wrapped worldvolume dimensions. The fivebrane is wrapped $N$ times on
the torus $E(\tau)$ in the M-theory spacetime with no branch-points or
other singularities. Thus the M5 world-volume has the form
$\R^{3,1}\times \Sigma$ where $\Sigma$ is an unbranched N-fold cover 
of $E(\tau)$. As the M5 winds $N$ times round the $x_{6}$ direction
and once round the $x_{10}$ direction the relevant $N$-fold covering 
has complex structure $\tilde{\tau}=N\tau$. 
The NS5 brane lifts to a second M5 brane located at
$z=0$ and infinitely extended in the $x_{4}$ and $x_{5}$ directions. 
In the case $N=2$, this configuration of two intersecting M5 branes 
precisely corresponds to the factorization curve described in the
previous section. As above, the Higgs branch corresponds to moving the 
two branes apart in their common transverse dimensions.    
 
Another interesting aspect of the results presented above 
was their invariance under 
$SL(2,{\bf Z})$ transformations acting on the complexified coupling 
constant $\tau$ and on the deformation parameter 
$\beta=-2\pi iRM$. In Section 5.4, we explained the significance of this
duality in the context of the $\beta$-deformed theory. It is also
of interest to understand this duality in the context of the 5d
theory. In the undeformed case, $M=0$, we are considering the
maximally supersymmetric $U(N)$ gauge theory on 
$\R^{3,1}\times \S^{1}$. At energies far below the compactification
scale $1/R$, this reduces to ${\cal N}=4$ SUSY Yang-Mills in four
dimensions with complexified coupling $\tau$, and the $SL(2,{\bf Z})$
in question is simply the usual S-duality of this theory. 
However this is not only a duality of the low-energy theory. In fact,
as we now review, it is an
exact duality of the full theory on $\R^{3,1}\times \S^{1}$. 

Recall that the maximally supersymmetric theory in five
dimensions is itself equivalent to a compactification of the 
$(2,0)$ superconformal theory which lives in six dimensions. 
Specifically the 5d theory with $M=0$ (and $\Theta=0$) corresponds to a
compactification of the $A_{N-1}$ $(2,0)$ theory\footnote{To produce a
  $U(N)$ gauge theory in five dimensions we should also include an
  additional free tensor multiplet in the six dimensional theory.} 
on $\R^{3,1}\times \S^{1}\times \tilde{\S}^{1}$. The 
first compact dimension with radius $R$ is already apparant in the 
5d theory. The second circle has radius set by the five-dimensional
gauge coupling: $\tilde{R}=G_{5}^{2}/8\pi^{2}$. From the
six-dimensional viewpoint, the electric-magnetic
duality transformation $\tau\rightarrow -1/\tau$ (with $\Theta=0$)
simply corresponds to an interchange of the two compact dimensions: 
$R\leftrightarrow \tilde{R}$. This implies the exact equivalence of
the 5d theory with coupling $G_{5}^{2}$ and 
radius of compactification $R$, with a dual theory with coupling 
$\tilde{G}_{5}^{2}=8\pi^{2}R$ and compactification radius 
$\tilde{R}$. More generally, we may introduce a four-dimensional
vacuum angle $\Theta$ by replacing $\S^{1}\times \tilde{\S}^{1}$ 
with a slanted torus of complex structure $\tau$. The 
$SL(2,{\bf Z})$ duality group acting on $\tau$ corresponds to the
diffeomorphism group of this torus. 
             
Our results, and those of \cite{Nek}, 
indicate that this duality extends to the 5d theory with a non-zero 
hyper-multiplet mass. In particular the dimensionless combination
$MR$, like the deformation parameter $\beta$, transforms with 
modular weight $(-1,0)$ under this duality. 
A related point is that $MR$ like $\beta$ has two periods. 
\EQ{
MR \rightarrow MR +i \ ,\qquad  MR \rightarrow MR +i\tau\ . 
\label{periodsmr}
}
The first period is already apparant in the classical theory and is a 
consequence of the identification 
of the twisted mass $\mu=2\pi R{\rm Im}[M]$ as a background Wilson
line. The second period (as noted in \cite{Nek}) is non-perturbative in the
coupling and is therefore invisible in the classical theory.    
It would be interesting to understand these features from the point   
of view of the six-dimensional $(2,0)$ theory.
 
One of the most interesting features of the $\beta$ deformed theory 
is the existence of branches of confining vacua corresponding to the
condensation of magnetic monopoles. The existence of these branches 
was first argued in \cite{nd1} as a consequence of S-duality 
acting on the Higgs branches of the theory and, in the previous
section, we checked this explicitly for gauge group $U(2)$.              
Having established the existence of both Higgs branches and S-duality
in the 5d theory it naturally follows that corresponding 
confining branches are also present in this theory. It is not hard to 
identify the massless states which condense on these branches. Recall 
that, in addition to elementary quanta, the 5d theory contains various 
solitonic states. In addition to ordinary 
four-dimensional magnetic monopoles which are independent of the
compact spatial coordinate, the theory contains solitons corresponding 
to Yang-Mills instantons on $\R^{3}\times \S^{1}$ treated as static
solutions of finite energy on $\R^{3,1}\times \S^{1}$. There are also 
boundstates of these objects which carry both instanton number and
magnetic charge. Each of these 
states saturates a BPS bound and their classical masses are 
determined by the central charge,          
\begin{equation}
\tilde{\cal Z}_{ab}=\frac{iq}{\tilde{R}}+ 
(\rho_{a}-\rho_{b})\frac{i\tilde{R}}{R} \pm M
\label{ccharge5d3}
\end{equation}     
where $\tilde{R}=G_{5}^{2}/R$ as above. Here $q$ corresponds to the 
instanton number while the contribution proportional to $\rho_{a}-\rho_{b}$ 
comes from states magnetically charged under 
the corresponding Cartan $U(1)$ subgroup. 
The central charge (\ref{ccharge5d3}) is evidently S-dual
to (\ref{ccharge5d})\footnote{We emphasize that both (\ref{ccharge5d3}) and 
(\ref{ccharge5d}) are classical formulae for the central charge. In a
theory with ${\cal N}=2$ supersymmetry, we generally expect that the
central charges and the BPS mass spectrum will recieve quantum corrections. 
Interestingly, in the present case, the classical formulae yield the 
exact results for the location of the Higgs and confining branches (ie
they agree with our earlier calculation based on the curve $\Sigma_{N}$). 
Thus, for these special values of the parameters and moduli, it seems
that the classical formulae (\ref{ccharge5d}) and (\ref{ccharge5d3})
are actually exact.}.    

The confining branch root occurs (for $\Theta=0$) 
at the point $MR=8\pi^{2} i/G_{5}^{2}N=iR/N\tilde{R}$. As at the Higgs
branch root, the eigenvalues of the adjoint scalar take the values 
$\rho_{a}=a i/NR$ for $a=1,2,\ldots,N$. Hence the central charge 
(\ref{ccharge5d3}) becomes 
\begin{equation}
\tilde{\cal Z}_{ab}=\frac{i}{N\tilde{R}}\left(qN+ a-b\pm 1\right)
\label{ccharge5d2}
\end{equation}   
Thus we find a total of $N$ massless ${\cal N}=2$ 
hypermultiplets coming from off-diagonal elements of 
$\tilde{\cal Z}$ with $a=b\pm 1$ mod $N$. Each of these fields 
carry opposite magnetic charges under a pair of adjacent $U(1)$'s 
in the Cartan subalgebra of $U(N)$. This precisely the same 
massless spectrum which appears at the root of the 
corresponding confining branch of the $\beta$ deformed theory.
 
The presence of the new branch can also be seen directly in the
M-theory construction of Section 5.3. Near the root of the branch, 
the curve factorizes into an M5 brane wrapped $N$ times on the torus 
$E(\tau)$ corresponding to the compact $x_{6}$ and $x_{10}$ directions 
and a single flat M5 brane extending in the $x_{4}$ and $x_{5}$
directions. This configuration is related to the configuration at the
Higgs branch root by a $6-10$ flip. Correspondingly the 
first M5 brane winds once around the $x_{6}$ direction and $N$ times
around the $x_{10}$ direction to give an $N$-fold cover of $E(\tau)$ 
with complex structure $\tilde{\tau}=\tau/N$. The existence of these 
branches of vacua was also noted in the string theory construction of 
\cite{RB}. 
 
The new branches of the 5d theory described above exhibit interesting
physics. We find the confinement of a $U(N)$ gauge group down
to its decoupled central $U(1)$ due to monopole condensation 
coexisting with unbroken ${\cal N}=2$ supersymmetry (in the four dimensional
sense). As in the $\beta$-deformed theory confinement occurs without a  
mass gap, although the interpretation of the resulting 
massless scalars is different. In the $\beta$-deformed theory, the
massless scalar fields could be interpreted as Goldstone bosons for
the spontaneously broken scale invariance and R-symmetry. In the 5d
case the massless scalars correspond to the traces of adjoint scalar
fields and are completely decoupled.  
 
The holomorphic equivalence described above only applies to the
Coulomb branch $\CB_{1}=\CB_{N,0,0}$ of the $\beta$-deformed theory
and the Higgs and confining branches which intersect it. It is
straightforward to extend these results to the generic Coulomb branch 
$\CB_{n_{1},n_{2},n_{3}}$ with $n_{1}+n_{2}+n_{3}=N$. This branch is 
described by a complex curve of the form $\Sigma_{n_{1}}\bigcup
\Sigma_{n_{2}}\bigcup\Sigma_{n_{3}}$ which is in turn appropriate to
govern the vacuum structure of the 5d theory with gauge group 
$U(n_{1})\times U(n_{2})\times U(n_{3})$.    

We can also extend our discussion to the $\beta$-deformed theory 
with gauge group $SU(N)$.  On the Coulomb sub-branch $\CB_1$ 
and at the level of the integrable system, the traceless constaint 
appropriate for the $SU(N)$ theory
corresponds to imposing the tracelessness of the Lax operator. This is
equivalent to imposing $H_1=0$.\footnote{Note that imposing $H_1=Nm$
  corrsponds to the $SU(N)$ theory with the mass term
  $m\Tr\,\Phi_2\Phi_3$ added to the superpotential}. The proof of this
simple condition is somewhat involved and we have relegated it to
Appendix C.
Notice that the constraint is different from the
constraint that gives the five-dimensional $SU(N)$ theory from the
five-dimensional $U(N)$ theory. In that case, the $U(1)$ factor
corresponds to the centre-of-mass motion of the integrable system, so
the constraint is $\sum_ap_a=0$, or $H_N=1$. This means that the
holomorphic equivalence of the two theories is not valid when the
gauge group is $SU(N)$. On one of the more general Coulomb
sub-branches one simply imposes $H_1$ in each of the three copies of
the RS integrable system.

In closing we note that the holomorphic equivalence described in this 
section suggests a direct way of realising the $\beta$-deformed theory 
on the world volume of Type IIA branes. The set up involves $N$ D4
branes wrapped around a compact dimension $x_{6}\sim x_{6}+2\pi R_{6}$
and extended in the $\{0,1,2,3\}$ directions. We will also include 
a single NS5 brane extended in the $\{0,1,2,3,4,5\}$ directions. 
We define a complex coordinate $U=x_{4}+ix_{5}$.  
So far this is simply a four dimensional version of the construction 
of subsection 5.2. 
The low energy theory on the branes (at energy scales much less than 
$1/R_{6}$) is just ${\cal N}=4$ SUSY Yang-Mills with gauge group 
$U(N)$. As in eqn (\ref{twist}) of 
subsection 5.2, an adjoint hypermultiplet mass can be
introduced by introducing an additive shift in $U$ on going around 
the $x_{6}$ circle. This leads to the standard brane construction of
the ${\cal N}=2^{*}$ theory in four dimensions given in 
\cite{witm}. Instead of doing this, we will introduce a 
{\em multiplicative} twist in $U$ via the identification.
\EQ{
 x_{6}\rightarrow x_{6}+2\pi R_{6}\ ,\qquad U \rightarrow \exp(i\beta)U\ .
 \label{twist3}
}    
It is straightforward to verify that the resulting spectrum of
stretched strings reproduces the classical spectrum of the 
$\beta$-deformed theory on its Coulomb branch $\CB_{1}$. One may 
also show that lifting this configuration to M-theory correctly 
reproduces the curve $\Sigma_{N}$ given in Section 5 above. 
It is tempting to conclude that the twist (\ref{twist3}) has the
effect introducing the $\beta$-deformation of the ${\cal N}=4$ theory 
on the worldvolume of the branes. 
However, this cannot be quite correct as 
the other Coulomb branches of the $\beta$-deformed theory are not visible 
in this brane construction. Equivalently, the construction only
produces the correct deformation of vacua in the appropriate
region of the Coulomb branch of the ${\cal N}=4$ theory.     
It would be interesting to understand this in more detail.

ND acknowledges useful discussions with Ofer Aharony. 
ND would like to thank the organizers of the `QCD and
String Theory' workshop at the ITP Santa Barbara where part of this
work was completed. TJH would like to thank all the organizers and 
participants of the 2nd Simons Workshop on Mathematical Physics in
Stony Brook for providing a 
stimulating atmosphere for completing this work.

\startappendix

\Appendix{Properties of the Curve}

To check 
property ${\bf (i)}$ we need the quasi-periodic properties of the function 
$\sigma(z)$:
\EQ{
\sigma(z+2\omega_{1})=
-\sigma(z)\exp\left(2(z+\omega_{1})\xi(\omega_{1})\right)\ ,\quad
 \sigma(z+2\omega_{2})=
-\sigma(z)\exp\left(2(z+\omega_{2})\xi(\omega_{2})\right)
}
and the relation $\omega_{2}\xi(\omega_{1})-\omega_{1}\xi(\omega_{2})
=i\pi/2$. Recall in our conventions $2\omega_1=2\pi i$ and
$2\omega_2=2\pi i\tau$. 
Using these relations we find that, 
\EQ{
L_{ab}(z+2\pi i)=\left(U_{1}L(z)U_{1}^{-1}\right)_{ab}\ , 
\qquad L_{ab}(z+2\pi i\tau)=  e^{i\beta}
\left(U_{2}L(z)U_{2}^{-1}\right)_{ab}\ , 
}
where the gauge transformations $U_{1}$ and $U_{2}$ are given by, 
\EQ{
\left(U_{1}\right)_{ab}=\exp\left(2x_{a}\xi(\pi i)\right)\ , 
\qquad
\left(U_{2}\right)_{ab}=\exp\left(2x_{a}\xi(\pi i\tau)\right) 
}
Using these relations in (\ref{spectral}) we can easily check that 
$F(z+2\pi i,x)=F(z,x)$ and $F(z+2\pi i\tau, e^{-i\beta}x)=F(z,x)$ 
as required by condition {\bf (i)}. 
 
To check condition ${\bf (ii)}$, we need to find the singularities 
of $F(z,x)$, which arise at singular points of the Lax matrix 
elements $L_{ab}(z)$ given in (\ref{lax}) above. The quasi-elliptic function 
$\sigma(z)$ behaves as $\sigma(z)\sim z+O(z^{5})$ near 
$z=0$ and has no other zeros or 
singularities in the period 
parallelogram. Thus the only singularity of $F(z,x)$ lies at $z=0$. Near 
$z=0$ we have, 
\EQ{
L_{ab}(z)\thicksim \frac1zL^{(-1)}_{ab}+ L^{(0)}_{ab}+{\cal O}(z)\ ,
}
where $L^{(-1)}_{ab}=i\rho_{a}$. 
As $L^{(-1)}$ is a projection operator onto the 
vector $(1,1,\ldots,1)$ we can easily change basis so that, 
\EQ{
\tilde{L}^{(-1)}_{ab}=\left(UL^{(-1)}U^{-1}\right)_{ab}=
i\tilde\rho_{a}\delta_{b1}
\label{newbasis}
}
for some element $U\in GL(N,{\bf C})$. We also define,
\begin{equation} 
\tilde{L}^{(0)}_{ab}=\left(UL^{(0)}U^{-1}\right)_{ab}\ .
\end{equation}
As the determinant defining $F(z,v)$ is invariant under this change of basis 
we can write, 
\SP{
F(z,x)& = {\rm det}\left(L(z)-x I_{(N)}\right)\thicksim {\rm det}\left(
\frac1z\tilde{L}^{(-1)}+\tilde{L}^{(0)}-xI_{(N)}\right)+{\cal O}(z) 
\\
&=\left|\begin{array}{cccc} 
\frac iz\tilde{Q}_{1}-x-\tilde{L}^{(0)}_{11} & \tilde{L}^{(0)}_{12} 
& \ldots \ldots & \tilde{L}^{(0)}_{1N} \\
i\tilde{Q}_{2}\frac{1}{z}+\tilde{L}^{(0)}_{21} & -x+\tilde{L}^{(0)}_{22} 
& \ldots \ldots & \tilde{L}^{(0)}_{2N} \\ 
\ldots & \ldots & \ldots\ldots & \ldots \\ 
\ldots & \ldots & \ldots\ldots & \ldots \\
i\tilde{Q}_{N}+\frac{1}{z}\tilde{L}^{(0)}_{N1} & \ldots & \ldots \ldots & 
\tilde{L}^{(0)}_{NN} \end{array} \right|  +{\cal O}(z)\ . 
\label{spectral2}
}
Evaluating the leading term in this determinant we find,
\begin{equation}
F(z,x) \thicksim \frac{i\tilde\rho_1f(x)}z+ O(z^0)
\end{equation}
where $f(x)$ is a polynomial in $x$ of order $N-1$. This verifies condition 
${\bf (ii)}$.

\Appendix{The Instanton Calculus}

In this appendix, 
we briefly describe how one can write down the
collective coordinate integrals over the moduli space of instantons
which determine the instanton contribution to the low-energy couplings
$\tau_{ab}$. We also sketch how one can show that the instanton
contribution to the couplings $\tau_{ab}=0$ when $a\in\II_i$,
$b\in\II_j$ with $i\neq j$. 

The instanton caluclus in the $\beta$-deformed theories can be deduced
from that of the $\N=4$ theory by taking a careful account of the
$\beta$ deformation. We will use the notation and results from the
review \cite{review}. At leading order in the semi-classical
expansion---which is the level required to calculate the instanton
contributions to $F$-term---the effect of the $\beta$ deformation is
to modify a subset of the Yukawa couplings. In order to describe the
deformation, we first relate the $\N=4$ notation of \cite{review} to
the $\N=1$ notation which is appropriate to the situation at hand.
In the $\N=4$ theory the relevant Yukawa couplings are of the form 
\EQ{
\text{Tr}(\lambda^{\alpha
  A}\bar\Sigma_{\hat aAB}[\varphi_{\hat a},\lambda^B_\alpha])\ .
\label{yuk}
}
Here, $\varphi_{\hat a}$ is an $SO(6)$ 
vector.\footnote{In order to avoid confusion with the gauge index, we
  indicate $SO(6)$ indices as $\hat a$.}
First of all, let us relate this to the language of $\N=1$.  
The three $\N=1$ chiral fields $\Phi_i=(\phi_i,\psi_{i\alpha})$, $i=1,2,3$,
are given by
\EQ{
\Phi_1=(-\varphi_5+i\varphi_6,\lambda^1_\alpha)\
,\quad\Phi_2=(\varphi_3-i\varphi_4,\lambda^2_\alpha)\ ,
\quad\Phi_3=(-\varphi_1+i\varphi_2,\lambda^3_\alpha)\ ,
\label{thh}
}
so note that $\psi_{i\alpha}\equiv\lambda^i_\alpha$,
while $\lambda^4_\alpha\equiv\lambda_\alpha$ is the gluino. The Yukawa
couplings \eqref{yuk} can then  be written in $\N=1$ language as
\EQ{
\sum_{ijk}\epsilon_{ijk}\psi_i^\alpha[\phi_j,\psi_{k\alpha}]+
\sum_i\psi_i^\alpha[\phi_i^\dagger,\lambda_\alpha]+\text{h.c.}\ .
}
The $\beta$-deformation then replaces the commutator in the first term
by the deformed commutator. In the language of $\N=4$, this can be
achieved by modifying the Clebsch-Gordon coeeficients in the following way:
\SP{
&\varphi_a\bar\Sigma_{\hat aAB}\to\varphi_a\bar\Sigma^{(\beta)}_{\hat aAB}\\
&=
\MAT{0 & e^{i\beta/2}(-\varphi_1+i\varphi_2) 
& e^{-i\beta/2}(-\varphi_3+i\varphi_4) & -\varphi_5-i\varphi_6\\
e^{-i\beta/2}(\varphi_1-i\varphi_2) 
& 0 & e^{i\beta/2}(-\varphi_5+i\varphi_6) & \varphi_3+i\varphi_4 \\
e^{i\beta/2}(\varphi_3-i\varphi_4) & e^{-i\beta/2}(\varphi_5-i\varphi_6)
 & 0 &-\varphi_1-i\varphi_2\\
\varphi_5+\varphi_6 & -\varphi_3-i\varphi_4 & \varphi_1+i\varphi_2&0}\ .
}

The fact that the $\beta$ deformation modifies the Yukawa couplings
does not affect the structure of the instanton (quasi-)zero
modes: at leading-order the gauge field and fermions take the same form in the
instanton backgound. The first effect is to change the
solution for the scalar fields at leading-order in the instanton
background but in a rather simple way. In the $\N=4$ theory, the solution
is given in Eq.~(4.64) of \cite{review}. One now simply replaces 
\EQ{
\bar\Sigma_{\hat aAB}\text{   by   }\bar\Sigma^{(\beta)}_{\hat aAB}\ .
\label{repl}
}
The same replacement in Eq.~(5.25) of \cite{review} gives $\tilde S^{(\beta)}$ 
the {\it instanton effective action\/} in the $\beta$ deformed
theory. In the $\N=4$ case, by introducing some auxiliary
variables one can relate the instanton collective coordinate system to
the theory of D-instantons inside D3-brane in Type IIB string
theory. We can capture the collective coordinate system of instantons
in the $\beta$-deformed theory by simply making the global replacement 
\eqref{repl}. The expression for the instanton action is then 
\EQ{
\tilde S^{(\beta)}=4\pi^2{\rm tr}\Big\{\big|w_\aD\chi_{\hat a}+
\varphi^0_{\hat a}w_\aD\big|^2-[\chi_{\hat
  a},a'_n]^2-\tfrac12\bar\Sigma^{(\beta)}_{\hat
  aBA}\bar\mu^A\varphi^0_{\hat a}\mu^B
\chi_{\hat a}+\bar\Sigma^{(\beta)}_{\hat aAB}(\bar\mu^A\mu^B+{\cal
  M}^{\prime A}{\cal M}^{\prime B})\chi_{\hat a}\Big\}+\tilde
S_\text{L.m.}\ ,
}
which replaces Eq.~(6.94) of \cite{review}. The Lagrange multipler
term imposes the bosonic and fermionic ADHM constraints:
\EQ{
\tilde S_\text{L.m.}=-4\pi^2\text{tr}\Big\{\bar\psi^\aD_A\big(\bar\mu^A
w_\aD+\bar w_\aD\mu^A+[{\cal M}^{\prime\alpha
  A},a'_{\alpha\aD}]\big)+D^c\big(\tau^{c\aD}_\bD(\bar w^\bD w_\aD+
\bar a^{\prime\bD\alpha}a'_{\alpha\aD})-\zeta^c\big)\Big\}\ .
}
In the above, the VEVs are the elements 
of the diagonal $N\times N$ matrix $\varphi^0_{\hat a}$ which are
given in terms of the $x_a$ by the correspondence \eqref{thh}. The
quantities $w_\aD$ and $\mu^A$ are $N\times k$ matrices while $\bar
w^\aD$ and $\bar\mu^A$ are $k\times N$ matrices (with $\bar
w^\aD\equiv(w_\aD)^\dagger$). The remaininf ones, ${\cal M}^{\prime A}_\alpha$,
$a'_{\alpha\aD}$, $D^c$, $\hat\chi_{\hat a}$ and $\bar\psi_A^\aD$ are all
$k\times k$ matrices.

Now we turn to the instanton contributions to the couplings
$\tau_{ab}$. These are given by the integrals over the instanton
moduli space in \eqref{insi}. 
In $\N=2$ theories the analogous integrals enjoy certain
localization properties
\cite{Hollowood:2002zv,Hollowood:2002ds}.\footnote{We remark that this
localization is more restricted than that used by Nekrasov
\cite{Nekrasov:2003af}, 
however, it has the advantage of easily extending to the $\N=1$ thoery
under discussion.}
We now argue that this localization
extends to the $\N=1$ theory. As in the $\N=2$ theories the argument rests on
the existence of a nilpotent fermionic symmetry $Q$. It the $\N=1$ theory
it is simply one of the two supersymmetries that are unbroken by the
instanton (taken with a $c$-number parameter). For example we can
choose $Q\equiv Q_1$. After the instanton action has been linearized
(see Section 6.5 of \cite{review}), it has the structure
\EQ{
\tilde S^{(\beta)}=Q\Xi+\Gamma\ ,
}
where $Q\Gamma=0$ so that $Q\tilde S^{(\beta)}=0$ (up to $U(k)$
transformations). Here,\footnote{Note that
  $\bar\Sigma^{(\beta)}_{\hat a4i}\equiv\bar\Sigma_{\hat a4i}=
-\bar\Sigma_{\hat ai4}$.} 
\EQ{
\Xi=-2i\pi^2
\bar\Sigma_{\hat a4i}\text{tr}\big(\bar\mu^i(w_1\chi_{\hat a}+\varphi^0_{\hat
  a}w_\aD)-(\chi_{\hat a}\bar w^2-\bar w^2\varphi^0_{\hat a})\mu^i\big)
}
and
\SP{
Q\Xi&=4\pi^2{\rm tr}\Big\{\big|w_\aD\chi_{\hat a}+
\varphi^0_{\hat a}w_\aD\big|^2+\tfrac12\bar\Sigma_{\hat
  a4i}(-\bar\mu^4\varphi^0_{\hat a}\mu^i+\bar\mu^i\varphi^0_{\hat
  a}\mu^4+\bar\mu^4\mu^i\chi_{\hat a}-\bar\mu^i\mu^4\chi_{\hat
  a})\\
&\qquad\qquad-\bar\psi^\aD_i(\bar\mu^iw_\aD+w_\aD\mu^i)\big\}\ .
}
It then follows that if we introduce a coupling $s$,
via $\tilde S^{(\beta)}\to s^{-1}Q\Xi+\Gamma$, then the resulting integrals
which give the coupling cannot depend on $s$. This follows from the
fact that the integrals are invariant under the supersymmetry $Q_1$. 
Taking $s\to\infty$, one can evaluate the integrals around the
zeros of $Q\Xi$. These are given by taking the $k\times k$ matrices 
$\chi_{\hat a}$ to be diagonal
with elements which equal one of the non-zero diagonal elements of the VEV
matrix $\varphi_{\hat a}$ (there are $N$ such elements $x_a$,
$a=1,\ldots,N$). So the number of critical points is equal to the
number of ways of distributing $k$ objects into $N$ sets. Suppose
$k\to k_1+\ldots+k_N$, where we allow $k_a=0$ for some values of
$a$. Each of the $k\times k$ 
matrix variables in the instanton calculus
has a block form which matches this partition: we denote the blocks
with the notation $[\cdots]_{ab}$. For example $[{\cal M}^{\prime
  A}]_{ab}$. At the critical points one can verify that all the matrix
variables are block diagonal. These collective coordinates describe an
instanton configuration which consist of $k_a$ abelian instantons in
the $a^\text{th}$ $U(1)$ subgroup of $U(N)$.\footnote{Instantons are
  non-trivial in an abelian theory once $|\zeta^c|>0$. This
  deformation is achieved by making the spacetime theory
  non-commutative. This deformation is not expected to affect the
  structure of the Coulomb branch \cite{review,Hollowood:2001ng}.}

One now expands around the critical point and integrates out fluctuations.
The leading order expression is then exact because
higher-order terms would depend non-trivially on powers of $s$. The
leading-order expression is itself independent of $s$ because of
cancellations between the bosonic and fermionic fluctuation
determinants. We are interested in the VEV dependence of the couplings
$\tau_{ab}$, such dependence arises when one integrates out the
off-diagonal fluctuations between each pair of blocks $a$ and $b$
($a\neq b)$. There are two generic situations. Firstly, when
$a\in\II_i$ and $b\in\II_j$ with $i\neq j$. In this case, taking for
example $i=1$ and $j=2$, the relevant
terms for the fermionic fluctuations 
in the instanton action are of the schematic form
\EQ{
[\bar\psi]_{ba}\MAT{0&0&-e^{-i\beta/2}x_b&-x_a^*\\
0&0&-e^{-\beta/2}x_a&x_b^*\\
e^{i\beta/2}x_b&e^{i\beta/2}x_a&0&0\\
x_a^*&-x_b^*&0&0}[\psi]_{ab}\ ,
\label{ffluc}
}
where $\psi$ and $\bar\psi$ are generic Grassmann collective
coordinates.\footnote{Either the pairs $(\bar\mu^A,\mu^A)$ or $({\cal
    M}^{\prime A}_1,{\cal M}^{\prime A}_2)$.} 
On integrating out these coordinates one gets determinants of the form
\EQ{
\big(|x_a|^2+|x_b|^2\big)^2
}
which does not depend at all on $\beta$. These determinants will then
cancel against bosonic determinants in the denominator.\footnote{This must
happen bacause the VEV dependence should disappear in the $\N=4$
theory at $\beta=0$.} The situations with $a,b\in\II_i$ is very
different. Taking $i=1$, for instance, 
the fermionic fluctuations 
in the instanton action are now of the schematic form
\EQ{
[\bar\psi]_{ba}\MAT{0&0&0&x_b^*-x_a^*\\
0&0&e^{i\beta/2}x_b-e^{-i\beta/2}x_a&0\\
0&e^{i\beta/2}x_a-e^{-i\beta/2}x_b&0&0\\
x_a^*-x_b^*&0&0&0}[\psi]_{ab}\ .
}
Now the determinant gives 
\EQ{
(e^{i\beta/2}x_a-e^{-i\beta/2}x_b)(e^{i\beta/2}x_b-e^{-i\beta/2}x_a)
(x_a^*-x_b^*)^2\ .
}
The compensating bosonic determinant is $\beta$-independent which
fixes the overall dependence on the pair $x_a$ and $x_b$ to be
\EQ{
\frac{(e^{i\beta/2}x_a-e^{-i\beta/2}x_b)(e^{-i\beta/2}x_a-e^{i\beta/2}x_b)}
{(x_a-x_b)^2}\ .
\label{poo}
}
This is R-symmetry invariant and holomorphic as required. Note that
these considerations match the result of perturbation theory as
described in Section 3.1.

In fact we can make a stronger statement about the couplings
$\tau_{ab}$ with $a\in\II_i$ and $b\in\II_j$ with $i\neq j$; namely
the couplings vanish. The reason depends on the behaviour of the 
insertions $\Xi_{a\alpha}$ in instanton integral. At leading around
the critical point $\Xi_{a\alpha}$ only depends on the collective
coordinates in the $a^\text{th}$ block. Hence, when  
$a\in\II_i$ and $b\in\II_j$ with $i\neq j$, the two insertions involve
the collective coordinates of two different blocks (in another words of
abelian instantons in two different $U(1)$ subgroups of the gauge
group). Inevitably this means that to get a non-zero answer in the
instanton integral entails going beyond leading order around the
critical point. Consequently the result must be a non-trivial function
of $s$ a dependence which is not allowed. Hence, we conclude that
\EQ{
\tau_{ab}=0\text{   when   }a\in\II_i\text{  and  }b\in\II_j\text{
  with  }i\neq j\ .
}

On the coulomb branch $\CB_i$ we have sketched above how the instanton
contributions can only depend on the VEVs through the functions
\eqref{poo}. This allows us to make a direct connection with the
Coulomb branch of the five-dimensional $\N=2^*$ theory compactified on
a circle. In this theory
the couplings (or pre-potential) are determined by instantons in much
the same way as above except that the collective coordinates can now
depend on $x^5$, the periodic coordinate. The instanton action is
precisely as for the $\N=4$ theory, but now there is an integral over
$t\equiv x^5$ and there is an additional mass term \cite{review}. Once again
the localizations arguments can be made. For us the interesting point
concerns the integrals over the fermionic fluctuations. The relevant
term in the instanton action mirrors \eqref{ffluc}, but with an integral
over $t$ and the addition of mass terms:
\EQ{
\int_0^R dt\,[\bar\psi]_{ba}(t)\MAT{0&0&0&\rho_b^*-\rho_a^*\\
0&0&\rho_b-\rho_a+M&0\\
0&\rho_a-\rho_b+M&0&0\\
\rho_a^*-\rho_b^*&0&0&0}[\psi]_{ab}(t)\ .
}
Here, $\rho_a$ are the VEVs of the adjoint scalar.
Now on integrating out the fluctuations one has to take account of all
the Kaluza-Klein modes around the circle. The end result is
\EQ{
\sinh \pi R(\rho_a-\rho_b+M)\sinh \pi R(\rho_b-\rho_a+M)
\sinh^2\pi R(\rho_a^*-\rho_b^*)\ .
}
Once again there must be a compensating determinant from the bosonic
fluctuations in order to cancel this when $M=0$ where the theory has
$\N=4$ supersymmetry. So the dependence on the VEVs is through 
\EQ{
\frac{\sinh 2\pi R(\rho_a-\rho_b+M)\sinh 2\pi R(\rho_a-\rho_b-M)}
{\sinh^22\pi R(\rho_a-\rho_b)}\ .
}
All the remaining parts of the calculation are identical the
four-dimensional case. Putting this together with the perturbative
contribution as explained in the text, it follows that the couplings of the
four-dimensional $\beta$-deformed theory on the Coulomb branch $\CB_i$ 
and five-dimensional $\N=2^*$ theory are formally related:
\EQ{
\tau_{ab}^{(5d)}(\rho_a,M)=\tau_{ab}\big(x_a=e^{2\pi
  R\rho_a},\beta=-2i\pi RM\big)\
.
}

\Appendix{$U(N)$ and $SU(N)$}

One can deduce the coulomb structure of the $SU(N)$ in the following way.
Firstly, at the level of the instanton calculus, one simply
imposes the tracelessness of the VEVs; in other words
\EQ{
\sum_{a\in\II_i}x_a=0\ ,
}
for $i=1,2,3$. This is because the instantons lie purely in the
non-abelian part of the gauge group. 
For example on the Coulomb sub-branch $\CB_1$ we have
$\sum_{a=1}^Nx_a=0$. 
Notice that in the compactified 
five-dimensional theory the constraint is different; namely
$\sum_{a=1}^N\rho_a=0$, i.e. $\prod_{a=1}^Nx_a=1$. So the holomorphic
equivalence of the $\beta$-deformed theory and the five-dimensional
theory is only true for $U(N)$ gauge groups.

One can show that the constraint $\sum_{a=1}^Nx_a=0$ becomes the
condition $H_1=0$ in the integrable system, or simply the
tracelessness of the Lax matrix. In order to see this we need to have
the mapping between the moduli $\{x_a\}$ and $\{H_a\}$. This can be
extracted from writing the curve of the five-dimensional theory in
terms of the moduli $\{\rho_a\}$:
\EQ{
\sum_{n=1}^\infty\frac1{n!}\Big(\frac{RM}i\Big)^n\partial_z^n
\theta_1(z/(2i)|\tau)\partial_u^n\prod_{a=1}^N\sinh(u-2\pi R\rho_a)\ .
}
Putting $e^{-u}=x$ and $RM=i\beta/(2\pi)$ gives the curve of the
$\beta$-deformed theory.
The first Hamiltonian $H_1$, or $\Tr\,L(z)$, is proportional to ratio
of the coefficients of the $x^{N-1}$ and $x^N$ terms. This gives 
\EQ{
H_1\propto\sum_{a=1}^Ne^{2\pi R\rho_a}=\sum_{a=1}^Nx_a\ ,
}
as expected.

\end{document}